\documentclass[preprint, 12pt]{elsarticle}

\usepackage{lipsum}

\usepackage{lineno}

\usepackage{amsmath,lscape}
\usepackage{amssymb}
\usepackage[utf8]{inputenc}

\usepackage{epsfig}
\usepackage{amsfonts}
\usepackage{fullpage}
\usepackage{rotating,array,booktabs} 
\usepackage{enumerate}
\usepackage[shortlabels]{enumitem}
\usepackage{wrapfig}
\usepackage{subfig}

\usepackage{multirow,multicol}
\usepackage{siunitx}
\usepackage{pdflscape}
\usepackage{afterpage}
\usepackage[nolist]{acronym}
\usepackage{longtable}

\usepackage{algpseudocode}
\usepackage{algorithm}

\usepackage{hyperref}
\hypersetup{
    unicode=true,          % non-Latin characters in Acrobat’s bookmarks
    pdftoolbar=true,        % show Acrobat’s toolbar?
    pdfmenubar=true,        % show Acrobat’s menu?
    pdffitwindow=false,     % window fit to page when opened
    pdfstartview={FitH},    % fits the width of the page to the window
    pdfnewwindow=true,      % links in new PDF window
    colorlinks=true,        % false: boxed links; true: colored links
    linkcolor= black,        % color of internal links (change box color with linkbordercolor)
    citecolor= black,        % color of links to bibliography
    filecolor= black,        % color of file links
    urlcolor= black          % color of external links
    breaklinks=true
}

\usepackage[table,xcdraw,dvipsnames]{xcolor}

\newcommand{\comment}[1]{#1}

\begin{document}
    
    \begin{frontmatter}
    % Carregar Informações do Artigo
        \title{Forecasting Brazilian and American COVID-19 cases based on artificial intelligence coupled with climatic exogenous variables}

\author[ad1]{Ramon Gomes da Silva}\corref{mycorrespondingauthor}
\ead{gomes.ramon@pucpr.edu.br}
\cortext[mycorrespondingauthor]{Corresponding author}

\author[ad1,ad2]{Matheus Henrique Dal Molin Ribeiro}
% \ead{matheus.dalmolinribeiro@gmail.com}

\author[ad3,ad4]{Viviana Cocco Mariani}
% \ead{viviana.mariani@pucpr.br}

\author[ad1,ad4]{Leandro dos Santos Coelho}
% \ead{leandro.coelho@pucpr.br}

\address[ad1]{Industrial \& Systems Engineering Graduate Program (PPGEPS), Pontifical Catholic University of Parana (PUCPR). 1155, Rua Imaculada Conceicao, Curitiba, PR, Brazil. 80215-901}

\address[ad2]{Department of Mathematics, Federal Technological University of Parana (UTFPR). Via do Conhecimento, KM 01 - Fraron,  Pato Branco, PR, Brazil. 85503--390}

\address[ad3]{Mechanical Engineering Graduate Program (PPGEM), Pontifical Catholic University of Parana (PUCPR). 1155, Rua Imaculada Conceicao, Curitiba, PR, Brazil. 80215-901}

\address[ad4]{Department of Electrical Engineering, Federal University of Parana (UFPR). 100, Avenida Coronel Francisco Heraclito dos Santos, Curitiba, PR, Brazil. 81530-000}

\journal{Chaos, Solitons \& Fractals}
    % Carregar resumo
        \begin{abstract}
The novel coronavirus disease (COVID-19) is a public health problem once according to the World Health Organization up to \comment{June 10th, 2020, more than 7.1 million people were infected, and more than 400 thousand have died worldwide.} In the current scenario, \comment{the} Brazil and the United States of America present a high daily incidence of new cases and deaths. Therefore, it is important to forecast the number of new cases in a time window of one week, once this can help the public \comment{health system} developing strategic planning to \comment{deals} with the COVID-19. The application of the forecasting artificial intelligence (AI) models has the potential \comment{of deal with} difficult dynamical behavior \comment{of time-series like of COVID-19.} In this paper, Bayesian regression neural network, cubist regression, $k$-nearest neighbors, quantile random forest, and support vector regression, are used stand-alone, \comment{and} coupled with the recent pre-processing variational mode decomposition (VMD) \comment{employed to decompose the time series into several intrinsic mode functions.} All AI techniques are evaluated in the task of time-series forecasting with one, three, and six-days-ahead the \comment{cumulative} COVID-19 cases in five Brazilian and American states, with \comment{a} high number of cases up to April 28th, 2020. Previous \comment{cumulative} COVID-19 cases and exogenous variables as daily temperature and precipitation were employed as inputs for all forecasting models. The models' effectiveness are evaluated based on the performance criteria. In general, the hybridization of VMD outperformed single \comment{forecasting} models regarding the accuracy, specifically when the horizon is six-days-ahead, the hybrid VMD--single models achieved better accuracy in 70\% of the cases. Regarding the exogenous variables, the importance ranking \comment{as predictor variables} is, from the upper to the lower, past cases, temperature, and precipitation. Therefore, due to the efficiency of evaluated models to \comment{forecasting cumulative COVID-19 cases} up to six-days-ahead, the adopted models can \comment{be recommended as a promising models for forecasting and be} used to assist in the development of public policies to mitigate the effects of COVID-19 outbreak.
\end{abstract}

    \end{frontmatter}
    
    % Enumerar linhas
    % \linenumbers

    % Corpo do Artigo
    \section{Introduction \label{INT}}

The new coronavirus disease (COVID-19) is a virus infectious disease induced by severe acute respiratory syndrome coronavirus 2 (SARS-CoV2). According to the World Health Organization (WHO), most of the population will mild to moderate respiratory illness and recover without requiring special treatment \cite{whoCOVID19}. However, several studies are being developed, and preliminary results indicated that people with underlying medical problems like cardiovascular disease, diabetes, chronic respiratory disease, obesity, and cancer are more likely to develop serious injuries \cite{BANSAL2020247,LAI2020105924,HUSSAIN2020108142,MOUJAESS2020102972,ABBAS2020100250,SU2020}. Also, the COVID-19 can cause extensive and multiple lung injuries \cite{Guan2020}, thus compromising the respiratory system of patients. In this context, the demand for devices that assist in the performance of breathing-related movements \comment{have increased.} 

Due to the serious damage caused by COVID-19, according to WHO, \comment{up to June 10th 2020, more than 7.1 million people were already infected, as well as more than 400 thousand people worldwide have now died with the coronavirus.} Indeed, considering the current scenario of the health system worldwide, the overcrowding could be observed \comment{in some countries, like Italy, Spain and perhaps Brazil. In Brazil context, believed that  the average of 3388 municipalities could have a significant deficit in hospital beds.} Especially, the deficit is projected to occur in \comment{Brazilian North and Northeast regions,} which means exceeding health care capacity due to the COVID-19 \cite{REQUIA2020139144}.  

Considering the importance of knowing the \comment{difficult} epidemiological scenario for COVID-19 on a short-term horizon, to mitigate the effects of this pandemic, the development of efficient \comment{and effective} forecasting models \comment{also has a positive impact on product reasonably accurate success rates forecasts the immediate future. Also, these models allow health managers to develop strategic planning and perform decision-making as assertively as possible. For this purpose, epidemiological models can be used, as it has been widely adopted in \cite{NDAIROU2020109846,BARMPARIS2020109842}. Alternatively, linear forecasting models \cite{ZHANG2020109829,CEYLAN2020138817,AHMAR2020138883}, artificial intelligence (AI) approaches \cite{RIBEIRO2020COVID,CHIMMULA2020109864}, as well as hybrid forecasting models \cite{CHAKRABORTY2020109850,SINGH2020109866} proved to be effective tools to forecast COVID-19 cases.} The advantages of AI approaches for time series forecasting lie in the flexibility of dealing with different kinds of response variables, as well as to the ability of these approaches to learning data dynamical behavior, complexity and accommodate nonlinearities, such as the observed in epidemiological data \cite{ribeiro2019Meningite}. Besides, hybrid methodologies allow us to combine several techniques such as pre-processing methods and single forecasting models.  

By the coupling of some methods, it is possible to use the specialty of each one to deal with different characteristics and therefore building an effective model. \comment{In context of the} preprocessing techniques, especially \comment{signal} decomposition methods, the variational mode decomposition (VMD) \cite{DRAGOMIRETSKIY2014} is an effective approach \comment{to decompose a dimensional signal into an ensemble of band-limited modes with specific bandwidth in a spectral domain} applied in several fields \cite{RODRIGUESMORENO2020112869,WU2019101657,LI2019240}, once can deal with nonlinearities, and non-stationarity inherent to time series. Considering the intrinsic mode function (IMF) obtained through VMD, it is hard to choose AI models to train and forecasting the VMD components. Therefore, based on this understanding, some models are coupled with VMD and are described in the following.

Due to the necessity of understanding the COVID-19 outbreak, \comment{and} the associated factors, or exogenous variables, some studies are being conducted considering the social environment, climatic variables, pollution, and population density \cite{PRATA2020138862,COCCIA2020138474,SHI2020138890,WU2020139051,AHMADI2020}. In this direction, in a general aspect, Sobral et al. \cite{SOBRAL2020138997} investigated the effects of climatic variables in COVID-19 spread for 166 countries. The authors argued that increasing the temperature reduced the COVID-19 cases, and precipitation also has a positive correlation with SARS-CoV2 cases. In the sequence, for Brazil, Auler et al. \cite{AULER2020139090} evaluated how meteorological conditions such as temperature, humidity, and rainfall can affect the spread of COVID-19 in five Brazilian cities. The authors concluded that higher mean temperatures and average relative humidity might support the COVID-19 transmission. Considering the United States of America (USA) weather aspects, especially for the New York state, Bashir et al. \cite{BASHIR2020138835} inferred that average and minimum temperature and air quality are significantly associated with the COVID-19 pandemic. All previously mentioned studies tried related the climatic variables with COVID-19 but \comment{in those papers} were not incorporated in time series models to forecasting COVID-19 cases. However, we think that incorporating the exogenous climatic variables in forecasting models can help to understand the data dynamic, and \comment{perhaps more} efficient forecasting models could be obtained \cite{ribeiro2020ensemble}.

In this respect, for forecasting \comment{of cumulative cases of COVID-19}, the objective of this paper is to explore and compare the predictive capacity of Bayesian regression neural network (BRNN), cubist regression (CUBIST), $k$-nearest neighbors (KNN), quantile random forest (QRF), and support vector regression (SVR) when are used stand-alone, \comment{and} a hybrid framework composed by VMD coupled with \comment{previously} mentioned models. \comment{In this study were used as datasets the number about the cumulative cases of COVID-19 from five Brazilian states} (Amazonas - AM, Ceara - CE, Pernambuco - PE, Rio de Janeiro - RJ, and Sao Paulo - SP), \comment{the first state from north region, the second and third states from northeast region, and the other two states from southeast region. Also were considered five  American states} (California - CA, Illinois - IL, Massachusetts - MA, New Jersey - NJ, and New York - NY)\comment{. The choice of these states was made through the largest} number of new cases of COVID-19 up to 28 April 2020.  

\comment{In the task of} forecasting horizons of the time series one, three, and six-days\comment{-ahead of cumulative COVID-19 cases} are adopted \comment{to evaluates the forecasting efficiency of the different models.} Additionally, previous COVID-19 cases, and exogenous variables such as daily temperature (maximum and minimum), and precipitation are employed as inputs for each evaluated model. The \comment{output}-of-sample forecasting accuracy of each model is compared by performance metrics such as the improvement percentage index (IP), symmetric mean absolute percentage error (sMAPE), and relative root mean squared error (RRMSE). Also, the importance of each input variable is presented for each country.

\comment{Forecasting models are impacted by the small dataset effect and the prediction of cases of COVID-19 a challenging task. The choice of the forecasting and pre-processing approaches is due to the fact that even that non-linear and AI models need large datasets to properly learn the data pattern, the use of exogenous variables (climate variables) and past values of the response variable overcomes this drawback.} 

\comment{VMD decomposes a time series into its intrinsic mode functions adaptively and non-recursively obtaining a set of sub-series with different features from low-frequency to high-frequency. The adoption of VMD with modes in conjunction with nonlinear prediction models of machine learning is a powerful framework to approach small datasets in forecasting task. In addition, BRNN and SVR approaches are capable of handling small samples, which makes them attractive for this study.}

The contributions of this paper can be summarized as follows: 

\begin{itemize}
    \item The first contribution is related to the proposal of two frameworks, non-decomposed and decomposed models, applied \comment{in the task of} forecasting the new \comment{cumulative cases of COVID-19} in five Brazilian and American states. It is expected that \comment{these} evaluated models can be used \comment{as} most accurate approaches to perform decision-making to structure the health system to avoid overcrowding in hospitals, \comment{and} preventing \comment{new} deaths. 
    
    \item The second contribution, we can highlight the use of a \comment{distinct} set of AI models based on machine learning approaches regarding learning structure, \comment{even} as the recent effective pre-processing VMD to forecasting the Brazilian and American COVID-19 new \comment{cumulative} cases. The forecasting models BRNN, CUBIST, KNN, QRF, SVR, and pre-processing VMD method were chosen once \comment{that} have reached success into several fields of regression and time series forecasting \cite{ribeiro2018multi,FERNANDEZDELGADO201911,ZUO2020124776,ZHU2019105739}; 
    
    \item Also, this paper evaluates AI models in a multi-day-ahead forecasting strategy coupled with climatic exogenous inputs. The range of the forecasting time horizon allows us to verify the effectiveness of the predicting models in different scenarios, associated with inputs such as previous COVID-19 \comment{cumulative} cases, temperature, and precipitation, allowing that the models achieve high forecasting accuracy. Finally, their results can help in planning actions to \comment{improve the health system to} contain \comment{the} COVID-19 \comment{deaths.}
\end{itemize}

The remainder of this paper is organized as follows: Section~\ref{material} a brief description of the dataset adopted in this paper is presented. The forecasting models applied in this study are described in Section~\ref{sub:BL}. Section~\ref{MET} details the procedures applied in the research methodology. Results obtained and related discussion about models’ forecasting performance are mentioned \comment{in} Section~\ref{RES}. Finally, Section~\ref{CONC} concludes this study with considerations and some directions for future research proposals.

    \section{Material and Methods \label{MAM}}

This section presents \comment{a} description of the material analyzed (Section \ref{material})\comment{,} as well as the model’s description applied in this paper (Section \ref{sub:BL}).

\subsection{Dataset Description \label{material}}

The collected dataset refers to the COVID-19 \comment{cumulative} cases that occurred in five states of \comment{the} Brazil and \comment{the} USA until April 20th, 2020. For the \comment{Brazilian} context, the dataset was collected from an API (Application Program Interface) \cite{brasilio} that retrieves the daily information about COVID-19 cases from all 27 Brazilian State Health Offices, \comment{assembles and makes} them publicly available. 
\comment{And for USA context, the dataset was collected from ``COVID-19 Data Repository'' on Github provided by the Center for Systems Science and Engineering (CSSE) at Johns Hopkins University \cite{csse}.}
The cumulative confirmed cases and deaths of each state, \comment{and} the period from the first and last reports, are illustrated in Table~\ref{tab:reports}. 

% TABLE - Confirmed cases data
\begin{table}[htb!]
\small
\centering
\caption{Summary of COVID-19 cases by country and state}
 \label{tab:reports}
\begin{tabular}{ccccccc}
\hline
Country &
  State &
  \begin{tabular}[c]{@{}c@{}}Number of\\ observed days\end{tabular} &
  \begin{tabular}[c]{@{}c@{}}Fist\\ reported\end{tabular} &
  \begin{tabular}[c]{@{}c@{}}Last\\ reported\end{tabular} &
  \begin{tabular}[c]{@{}c@{}}Cumulative\\ cases\end{tabular} &
  \begin{tabular}[c]{@{}c@{}}Cumulative\\ deaths\end{tabular} \\ \hline
\multirow{5}{*}{Brazil} & AM & 47 & 13/03/2020 & 28/04/2020 & 4337   & 351   \\
                        & CE & 44 & 16/03/2020 & 28/04/2020 & 6985   & 403   \\
                        & PE & 48 & 12/03/2020 & 28/04/2020 & 5724   & 508   \\
                        & RJ & 55 & 05/03/2020 & 28/04/2020 & 8504   & 738   \\
                        & SP & 64 & 25/02/2020 & 28/04/2020 & 24041  & 2049  \\ \hline
\multirow{5}{*}{USA}    & CA & 94 & 26/01/2020 & 28/04/2020 & 46164  & 1864  \\
                        & IL & 96 & 24/01/2020 & 28/04/2020 & 48102  & 2125  \\
                        & MA & 87 & 01/02/2020 & 27/04/2020 & 56462  & 3003  \\
                        & NJ & 55 & 05/03/2020 & 28/04/2020 & 113856 & 6442  \\
                        & NY & 58 & 02/03/2020 & 28/04/2020 & 295106 & 22912 \\ \hline
\end{tabular}
\end{table}

The climatic exogenous variables were retrieved from the ``\textit{Instituto Nacional de Meteorologia}'' (INMET) \cite{INMET} for data from Brazil\comment{, while the} USA climate dataset \comment{were taken into a count} from \comment{the} daily global historical climatology network \comment{that} was retrieved from the National Centers for Environmental Information (NCEI) from the National Oceanic and Atmospheric Administration \cite{NOAA}, by using \texttt{rnoaa} package \cite{RNOAA}. For each state, considering the daily available information, minimum and maximum temperature (\textit{ºC}), and precipitation (\textit{mm}) were select as climatic exogenous inputs \comment{to each forecasting model applied in} this study. The measurement period of each state is variable, \comment{this is due} because the record of the first case of the disease may differ from state to state. The summary of the climatic variables used is described in Table \ref{tab:climatic}.  

\begin{table}[htb!]
\scriptsize
\centering
\caption{Descriptive measures for climatic variables by country and state}
\label{tab:climatic}
\begin{tabular}{cclcccc}
\hline
Country & State & Variable & Minimum & Median & Mean & Maximum \\ \hline
\multirow{15}{*}{Brazil} & \multirow{3}{*}{AM} & Minimum temperature (\textit{ºC}) & 24.76 & 26.20 & 26.36 & 28.28 \\
 &  & Maximum temperature (\textit{ºC}) & 25.29 & 27.05 & 27.24 & 29.55 \\
 &  & Precipitation (\textit{mm}) & 0.00 & 0.11 & 0.33 & 2.40 \\ \cline{2-7}
 & \multirow{3}{*}{CE} & Minimum temperature (\textit{ºC}) & 25.14 & 26.62 & 26.60 & 27.90 \\
 &  & Maximum temperature (\textit{ºC}) & 25.91 & 27.73 & 27.68 & 28.99 \\
 &  & Precipitation (\textit{mm}) & 0.00 & 0.12 & 0.25 & 1.31 \\ \cline{2-7}
 & \multirow{3}{*}{PE} & Minimum temperature (\textit{ºC}) & 23.36 & 25.18 & 25.05 & 26.74 \\
 &  & Maximum temperature (\textit{ºC}) & 24.27 & 26.30 & 26.10 & 27.96 \\
 &  & Precipitation (\textit{mm}) & 0.00 & 0.14 & 0.23 & 1.33 \\ \cline{2-7}
 & \multirow{3}{*}{RJ} & Minimum temperature (\textit{ºC}) & 19.07 & 21.23 & 21.57 & 25.33 \\
 &  & Maximum temperature (\textit{ºC}) & 19.69 & 22.16 & 22.56 & 26.49 \\
 &  & Precipitation (\textit{mm}) & 0.00 & 0.03 & 0.13 & 1.32 \\ \cline{2-7}
 & \multirow{3}{*}{SP} & Minimum temperature (\textit{ºC}) & 17.60 & 19.99 & 20.09 & 23.40 \\
 &  & Maximum temperature (\textit{ºC}) & 18.76 & 21.11 & 21.37 & 25.03 \\
 &  & Precipitation (\textit{mm}) & 0.00 & 0.00 & 0.12 & 1.19 \\ \hline
\multirow{15}{*}{USA} & \multirow{3}{*}{CA} & Minimum temperature (\textit{ºC}) & -1.76 & 4.90 & 4.91 & 11.92 \\
 &  & Maximum temperature (\textit{ºC}) & 10.60 & 18.33 & 18.54 & 28.59 \\
 &  & Precipitation (\textit{mm}) & 0.01 & 4.66 & 20.98 & 162.67 \\ \cline{2-7}
 & \multirow{3}{*}{IL} & Minimum temperature (\textit{ºC}) & -19.42 & -0.42 & -0.75 & 14.63 \\
 &  & Maximum temperature (\textit{ºC}) & -6.39 & 8.40 & 8.99 & 26.06 \\
 &  & Precipitation (\textit{mm}) & 0.00 & 4.29 & 23.30 & 196.47 \\ \cline{2-7}
 & \multirow{3}{*}{MA} & Minimum temperature (\textit{ºC}) & -14.75 & -0.86 & -1.76 & 5.39 \\
 &  & Maximum temperature (\textit{ºC}) & -2.77 & 8.32 & 8.16 & 18.70 \\
 &  & Precipitation (\textit{mm}) & 0.00 & 6.84 & 34.66 & 320.86 \\ \cline{2-7}
 & \multirow{3}{*}{NJ} & Minimum temperature (\textit{ºC}) & -11.80 & 1.60 & 0.96 & 7.27 \\
 &  & Maximum temperature (\textit{ºC}) & 0.54 & 10.78 & 11.22 & 22.02 \\
 &  & Precipitation (\textit{mm}) & 0.00 & 5.44 & 31.53 & 274.04 \\ \cline{2-7}
 & \multirow{3}{*}{NY} & Minimum temperature (\textit{ºC}) & -20.48 & -2.61 & -3.75 & 4.60 \\
 &  & Maximum temperature (\textit{ºC}) & -7.98 & 5.62 & 6.23 & 17.88 \\
 &  & Precipitation (\textit{mm}) & 0.00 & 9.94 & 27.61 & 167.94 \\ \hline
\end{tabular}%
\end{table}

\begin{figure}[htb!]
    \centering
    \subfloat[\comment{Brazil} \label{subfig:heatmap_BRA}]{
    \includegraphics[width=.385\linewidth]{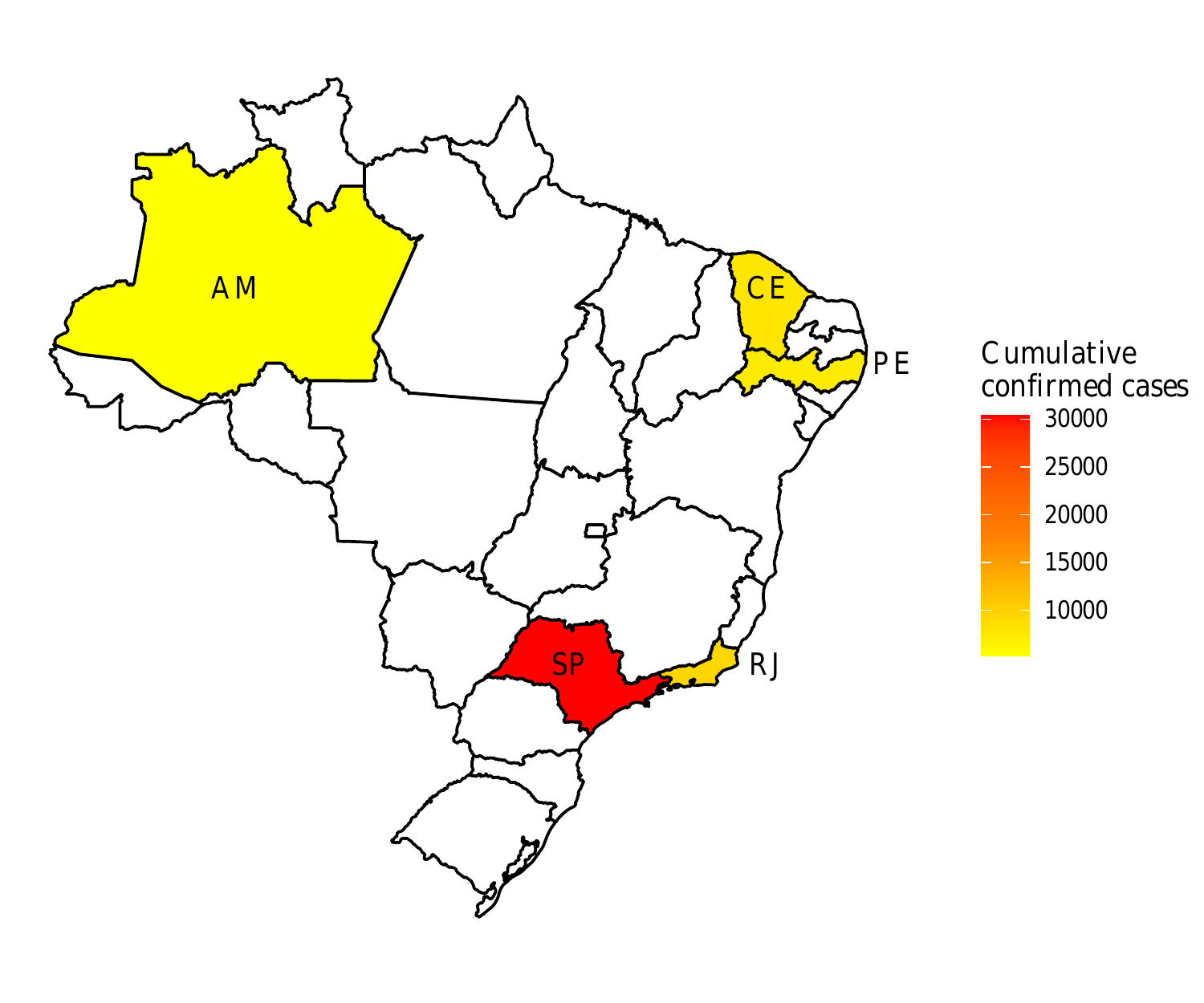}
    }
    \subfloat[USA \label{subfig:heatmap_USA}]{
    \includegraphics[width=.575\linewidth]{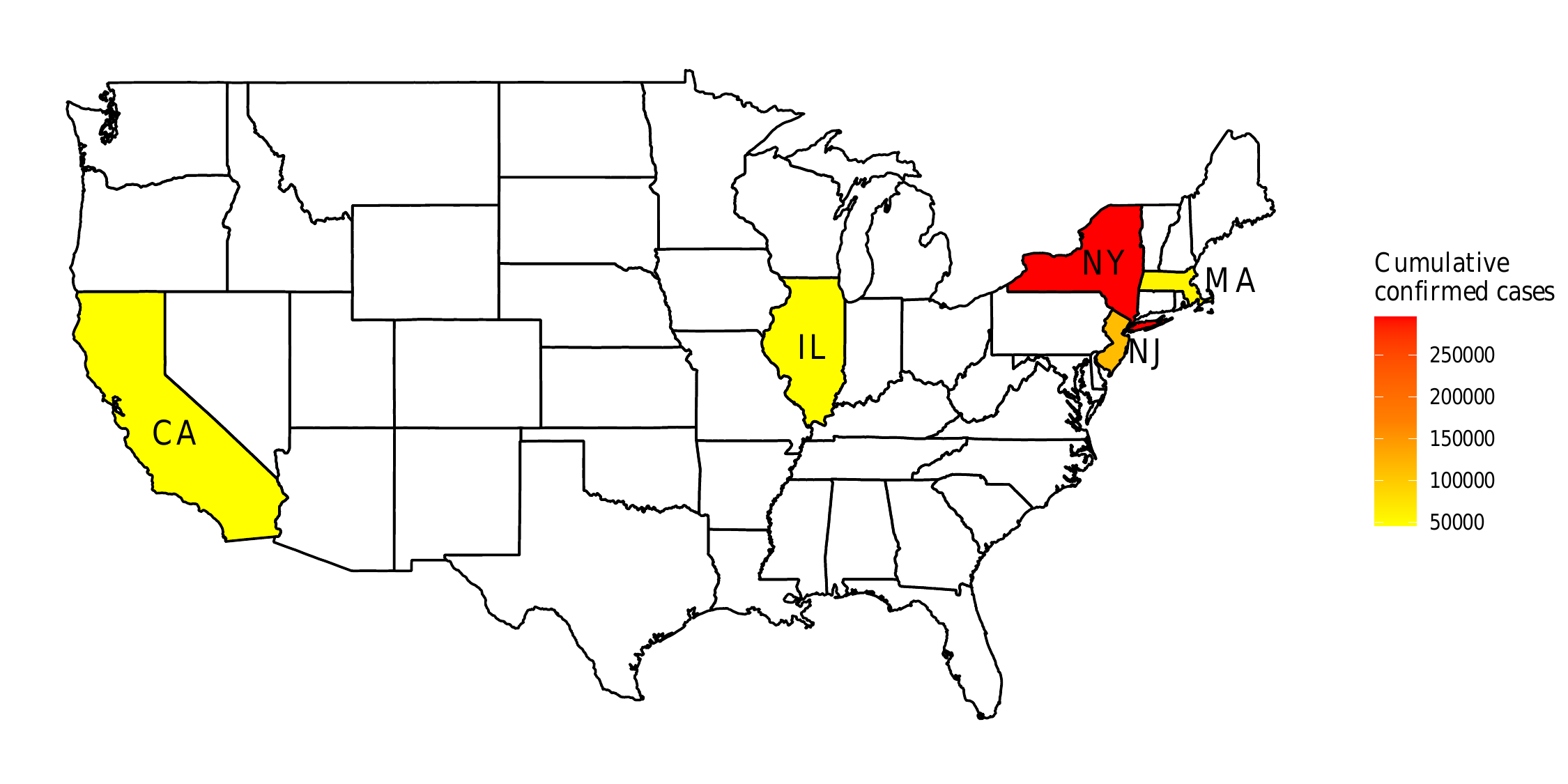}
    }
    \caption{Heatmap of the cumulative confirmed cases to five states from Brazil and USA.}
    \label{fig:heatmap}
\end{figure}

The heat-map of the cumulative confirmed cases from \comment{the} Brazil and the USA in each of the five states analyzed \comment{are} presented in Figure~\ref{fig:heatmap}. In that figure can be seen that the states with the highest number of COVID-19 \comment{cumulative} cases are SP and NY, respectively, in Brazil and the USA\comment{, the states with the highest demographic index in both countries.}

% FIGURE - Heatmap

\subsection{Methodologies \label{sub:BL}}

This section presents a summary of each model employed in the data analysis. 

\begin{itemize}
    \item BRNN is a kind of feedforward neural network, a two-layer neural network, composed by one input and one hidden layer, which uses the Bayesian methods, such as empirical Bayes, for parameter estimation, to avoid overfitting \cite{MacKay1992}. In the BRNN formulation, the variances are regularization parameters, in which the trade-off between goodness-of-fit and smoothing can be controlled. Also, in this approach the method of \cite{Nguyen} is used to assign initial weights of neural network and the Gauss-Newton training algorithm to perform the optimization. For the datasets evaluated in this paper, the BRNN becomes attractive once \comment{it} can deal with small samples, as well as \comment{it has} a lower computational cost.  

    \item CUBIST is a rule-based algorithm used to build forecasting models (in the time series field) based on the analysis of input data \cite{quinlan1992learning}. It estimates the target values by establishing regression models with one or more rules (committee/ensemble of rules) based on the input set. These rules are employed based on a combination of conditions with a linear function (in general linear regression). When the rule satisfies all conditions defined in the learning process, this approach can execute multiple rules once and find different linear functions suitable to forecast COVID-19 cases. However, if the standard deviation reduction value is smaller or equal to the expected error for sub-tree, some leaves are pruned to avoid overfitting \cite{RIBEIRO2020COVID}. 

    \item KNN is an instance-based learner model designed to solve classification and regression problems \cite{Aha1991}. In fact, in the time series context, the KNN searches \textit{k} nearest past similar values in the input set (past COVID-19 values, and climatic variables), in which these \textit{k} values are namely nearest neighbors. In this context, to find the nearest values, a similarity measure is adopted. The \textit{k}-nearest neighbors are those that similarity measure between past cases and new cases is the smallest. Considering that the set of \textit{k}-nearest neighbors are defined, the forecasting of new COVID-19 cases is obtained through of average of past similar values. In contrast to the simplicity of this supervised learning, the computational cost may be a disadvantage \cite{ribeiro2020ensemble}.

    \item QRF approach is an extension of \comment{the} random forests (RF) ensemble model \cite{Breiman2001}. It provides information about the full conditional distribution of the response variable, not only about the conditional mean. In this approach, the use of conditional quantile is to enhance the RF performance, which makes this a consistent approach \cite{meinshausen2006}. The main assumption about QRF lies in that weighted observations can be used for estimating the conditional mean \cite{Vaysse2017}. Additionally, while the RF approach keeps in the results information as regards the average cases of COVID-19 of the leaves, the QRF keeps all COVID-19 cases contained in the leaves. 

    \item SVR \comment{is a type support vector machine that} consists in determining support vectors close to a hyperplane, which maximizes the margin between two-point classes obtained from the difference between the target value and a threshold. To deal with non-linear problems SVR takes into account kernel functions, which calculates the similarity between two observations through the inner product. In this paper, the linear kernel is adopted. The main advantages of the use of SVR lie in its capacity to capture the predictor non-linearity and then use it to improve the forecasting cases. Also, it is advantageous to employ to forecast COVID-19 \comment{cumulative} cases, once the samples are small \cite{Drucker1997,RIBEIRO2020COVID}.

    \item VMD is a pre-processing technique in the field of decomposition approaches, which decomposes a time series into a finite and predefined $k$ number of Intrinsic Mode Functions (IMF) or mode functions. In a general way, VMD reproduces the decomposed signal with different sparsity properties \cite{DRAGOMIRETSKIY2014}. There are three main concepts related to VMD, which are Wiener filtering, Hilbert transform and analytic signal, and frequency mixing and heterodyne demodulation. Sparsity prior of each mode is chosen as bandwidth in the spectral domain and can be accessed by the following scheme for each model: (i) compute associated analytic signal utilizing the Hilbert transform to obtain a unilateral frequency spectrum; (ii) shift frequency spectrum of mode to baseband by mixing the exponential tune to the respective estimated center frequency; and (iii) the bandwidth estimated through the Gaussian smoothness of the demodulated signal \cite{RODRIGUESMORENO2020112869}. 
\end{itemize}

    \section{Proposed forecasting framework \label{MET}}

This section describes the main steps in the data analysis adopted by BRNN, CUBIST, KNN, QRF, SVR, and VMD based models. 

    \textbf{Step 1}: \label{step1} First, the dataset output variables are decomposed into five IMFs by performing VMD. The lag equal 2 was chosen by grid-search, applied on the IMFs creating four inputs from the lags, and applied on the exogenous inputs as well. Further, the new data is split into training and test sets. The test set consists of the last six observations and the training set defined by the remaining samples. In the training state, leave one-out-cross-validation with time slice was adopted, such as developed by \cite{ribeiro2020ensemble}.
    
    \textbf{Step 2}: \label{step2} Each IMF is trained with each model described in Section~\ref{sub:BL} using time-slice validation approach. Next, the IMF predictions were reconstructed by a simple summation-grouping model, in other words, the IMF is trained by the same model and is summed. Then, five predictions outputs were generated named VMD--BRNN, VMD--CUBIST, VMD--KNN, VMD--QRF, and VMD--SVR.
    
    \textbf{Step 3}: \label{step3} A recursive strategy is employed to develop multi-days-ahead COVID-19 cases forecasting \cite{RIBEIRO2020COVID}. Regarding this, one model is fitted for one-day-ahead forecasting, then the recursive strategy uses this forecasting result as an input for the same model to forecast the next step, continuing until the desirable forecasting horizon. In this study, the aim is to obtain the cases up to $H$ next days, especially up to 1 (ODA, one-day-ahead), 3 (TDA, three-days-ahead), and 6-days-ahead (SDA, six-days-ahead), respectively. The following structures are considered,
    \begin{equation}
        \hat{y}_{(t+h)} =
        \begin{cases}
        \hat{f}\left\{y_{(t+h-1)}, \,y_{(t+h-2)},\,\mathbf{X}_{(t+h-1)}\right\} & \text{if } h = 1, \\
        \hat{f}\left\{\hat{y}_{(t+h-1)},\,\hat{y}_{(t+h-2)},\, \mathbf{X}_{(t+h-3)}\right\} & \text{if } h = 3, \\
        \hat{f}\left\{\hat{y}_{(t+h-1)},\, \hat{y}_{(t+h-2)},\, \mathbf{X}_{(t+h-6)}\right\} & \text{if } h = 6, \\
        \end{cases}
    \end{equation}
    where $\hat{f}$ is a function that maps the cumulative COVID-19 cases, $\hat{y}(t+h)$ is the forecast of cumulative cases in horizon $h=$1, 3 and 6, $y(t+h-1)$, ${y}(t+h-2)$ are the previous observed, $\hat{y}(t+h-1)$, $\hat{y}(t+h-2)$ are the predicted cumulative cases, $\mathbf{X}(t+h-n_x)$ is the exogenous inputs vector at the maximum lag of inputs ($n_x = 1$ if $h = 1$, $n_x = 3$ if $h = 3$, and $n_x = 6$ if $h = 6$).The analyses are developed using \texttt{R} software \cite{R}. All \comment{hyperparameters} employed in this study are presented in Tables \ref{tab:hyper} and \ref{tab:hyper2} in \ref{apendixB}.  
    
    \textbf{Step 4}: To evaluate the effectiveness of adopted models, from obtained forecasts out-of-sample (test set), performance IP  \eqref{eq:criteria1}, sMAPE \eqref{eq:criteria2}, and RRMSE \eqref{eq:criteria3} criteria are computed as
    \begin{equation}
        \begin{aligned}
            \operatorname{IP}&=&100 \times \frac{\displaystyle M_c-M_b}{M_c},
        \label{eq:criteria1}
        \end{aligned}
    \end{equation}
    \begin{equation}
        \begin{aligned}
            \operatorname{sMAPE}=\frac{2}{n}\displaystyle \sum_{i=1}^{n} \frac{\left|y_i-\hat{y}_i\right|}{\left|y_i\right|+\left|\hat{y}_i\right|},
                  \label{eq:criteria2}
        \end{aligned}
    \end{equation}  
    \begin{equation}
        \begin{aligned}
            \operatorname{RRMSE}=\frac{\sqrt{\frac{1}{n} \displaystyle\sum_{i=1}^{n}\left(y_{i}-\hat{y}_{i}\right)^{2}}}{\frac{1}{n} \displaystyle\sum_{i=1}^{n} y_{i}},      \label{eq:criteria3}
        \end{aligned}
    \end{equation}  
    where $n$ is the number of observation, $y_i$ and $\hat{y}_{i}$ are the \textit{i}-th observed and predicted values, respectively. Also, the $M_c$ and $M_b$ represent the performance measure of compared and best models, respectively.

\begin{figure}[htb!]
    \centering
    \includegraphics[width=0.9\linewidth]{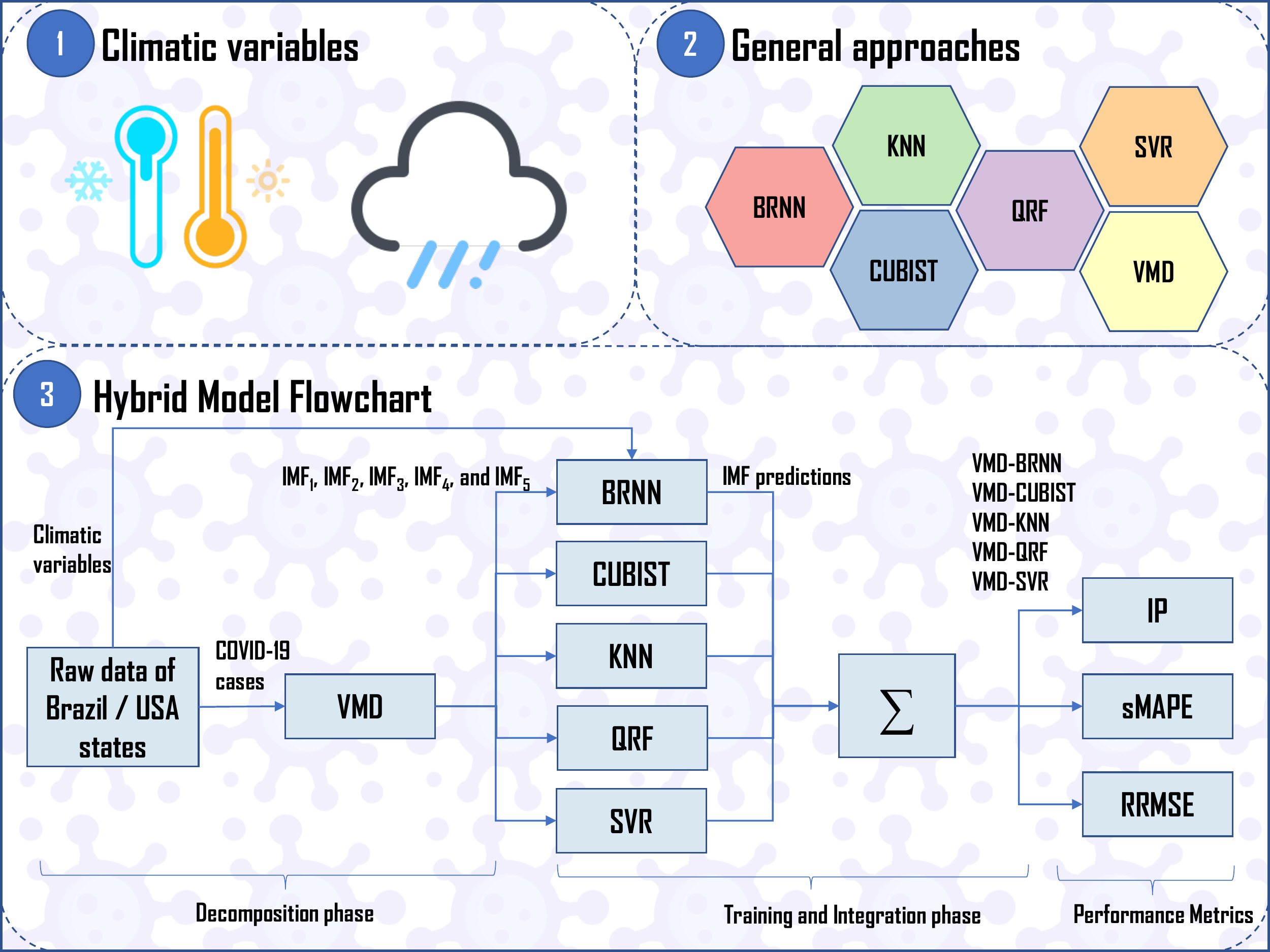}
    \caption{Proposed forecasting framework}
    \label{fig:flowchart}
\end{figure}

Figure \ref{fig:flowchart} presents the proposed forecasting framework.

% FIGURE - Diagram

    \section{Results \label{RES}}

This section describes the results of the developed experiments in forecasting out-of-sample (test set). First, Section \ref{sec:E1} compares the results of evaluated models over ten datasets and three forecasting horizons adopted. In Tables \ref{tab:performancemeasure} and \ref{tab:performancemeasure2} in \ref{apendixA}, the best results regarding accuracy are presented in bold. Additionally, Figures~\ref{fig:PObrazil} and~\ref{fig:POusa} illustrate the relation between observed and predicted values achieved by models with the best set of performance measures depicted in Tables \ref{tab:performancemeasure} and \ref{tab:performancemeasure2}, as well as box-plots for out-of-sample errors, are illustrated in Figure \ref{fig:error}. Also, Figure~\ref{fig:importance} illustrates the variable importance of each input (both lags and exogenous inputs) used in the models' predictions.

\subsection{Performance measures for compared models \label{sec:E1}}

In this section, the main results achieved by the best model regarding sMAPE and RRMSE criteria are presented for short-term forecasting multi-days-ahead of cumulative cases of COVID-19 from five Brazilian and five American states. 

Firstly, considering the results for the Brazil context, the main results are highlighted as follows.

\begin{itemize}
    \item AM: In this state, VMD--BRNN could be considered to forecasting COVID-19 cases, once the model outperformed all the single and VMD models in both performance criteria in all forecasting horizons. The improvement in the sMAPE achieved by VMD--BRNN ranges between 39.47\% - 96-06\%, 55.97\% - 94.88\%, and 67.41\% - 94.25\%, for ODA, TDA, and SDA horizon respectively. Regarding RRMSE analysis, the improvement ranges between 9.86\% - 94.81\%, 33.44\% - 93.29\%, and 56.66\% - 93.89\%, respectively.
    
    \item CE, RJ, and SP: For these states, in all forecasting horizons, the VMD--CUBIST approach achieved better accuracy than other models, for both sMAPE and RRMSE criteria in the multi-days-ahead forecasting task of the confirmed number of COVID-19. In fact, the improvement in sMAPE is ranged in 8.67\% - 96.57\%, 12.15\% - 97.78\%, and 59.37\% - 97.09\%, respectively, in ODA, TDA, and SDA forecasting horizons. Moreover, the improvement in RRMSE is ranged in 12.41\% - 97.32\%, 2.61\% - 98.29\%, and 49.99\% - 97.95\%, respectively.

    \item PE: In this state, CUBIST and SVR present better performance to forecasting COVID-19 cases. For ODA and TDA, CUBIST outperforms models, while for SDA the SVR achieves better accuracy regarding sMAPE and RRMSE than others. The improvement in the sMAPE for ODA and TDA achieved by CUBIST ranges between 6.81\% – 97.93\%, and 24.94\% – 98.23\%, respectively. For SDA, SVR outperforms other models, and this criterion is reduced in the range of 49.36\% - 98.27\%. Moreover, the same behavior is observed when the improvement in \comment{the} RRMSE criterion is obtained.
\end{itemize}

\textbf{Remark:} In this experiment, regarding the Brazilian states, 150 scenarios (5 datasets, 3 forecasting horizons, and 10 models) were evaluated for the task of forecasting cumulative COVID-19 cases. In an overview, the best models for each state, obtained sMAPE ranged between 1.14\% - 3.05\%, 1.06\% - 2.79\%, and 1.05\% - 3.03\% for ODA, TDA, and SDA forecasting, respectively. In the Brazilian context, the ranking of the model in all scenarios is VMD--CUBIST, VMD--BRNN, SVR, CUBIST, VMD--SVR, BRNN, VMD--QRF, QRF, VMD--KNN, and KNN. From a broader perspective, the efficiency of the VMD models is due to the capability of the approach to deal with non-linearity and non-stationarity of the data. Moreover, the efficiency of the CUBIST is due mainly to its ensemble learning of rules, \comment{in} which the approach \comment{takes advantage} of each rule based on the input set. On the other hand, the difficulty of the KNN model to forecasting cumulative COVID-19 cases could be attributed to the fact that this approach requires more observations to effectively learn the data pattern, once the forecasting is obtained by an average of past similar values. 

In the next, considering the results for the USA context, the main results are highlighted as follows.

\begin{itemize}
    \item CA: In CA state, BRNN outperformed other models, in all forecasting horizons, for both sMAPE and RRMSE criteria. In this aspect, the improvement in sMAPE ranges between 29.98\% - 97.86\%, 4.64\% - 97.71\%, and 48.56\% - 97.99\%, for ODA, TDA, and SDA, respectively. Regarding RRMSE, the improvement ranges in 24.00\% - 97.67\%, 6.57\% - 97.78\%, and 48.62\% - 98.11\%, respectively.
    
    \item IL, MA, and NJ: For both performance criteria, CUBIST outperformed other models in ODA, for IL and NJ states, and TDA, for IL. BRNN presented better accuracy than other models, for MA state in ODA and TDA. Moreover, VMD--CUBIST outperformed other models in SDA for these three states. In fact, the improvement in sMAPE is ranged in 6.63\% - 98.76\%, 31.89\% - 98.09\%, and 3.76\% - 97.98\%, respectively, in ODA, TDA, and SDA forecasting horizons. Moreover, regarding the RRMSE, the improvement ranges between 7.54\% - 98.48\%, 0.83\% - 98.25\%, and 3.25\% - 98.11\%, respectively.
    
    \item NY: For NY state, in both performance criteria, VMD--CUBIST presented better accuracy than other model in ODA forecasting, while SVR outperformed the other models in TDA and SDA forecasting. Regarding sMAPE, the improvement ranges 17.86\% - 95.44\%, 16.12\% - 95.69\%, and 42.39\% - 92.71\%, for ODA, SDA, and TDA, respectively. For RRMSE, the improvement ranges 25.78\% - 96.09\%, 7.78\% - 95.43\%, and 43.76\% - 93.45\%, respectively.
\end{itemize}

\textbf{Remark:} In this experiment, regarding the American states, 150 scenarios (5 datasets, 3 forecasting horizons, and 10 models) were evaluated for the task of forecasting cumulative COVID-19 cases. In an overview, the best models for each state, obtained sMAPE ranged between 0.54\% - 1.90\%, 0.55\% - 1.59\%, and 0.62\% - 3.08\% for ODA, TDA, and SDA forecasting, respectively. In the American context, the ranking of the models in all scenarios is VMD--CUBIST, BRNN, CUBIST, SVR, VMD--BRNN, VMD--SVR, VMD--QRF, QRF, KNN, and VMD--KNN. The same behavior presented in Brazilian cases is presented in the American, which the VMD--CUBIST in overall had better average performance compared to the other models.

\begin{figure}[htb!]
    \centering
    \subfloat[AM \label{subfig:AM}]{
    \includegraphics[width=0.43\linewidth]{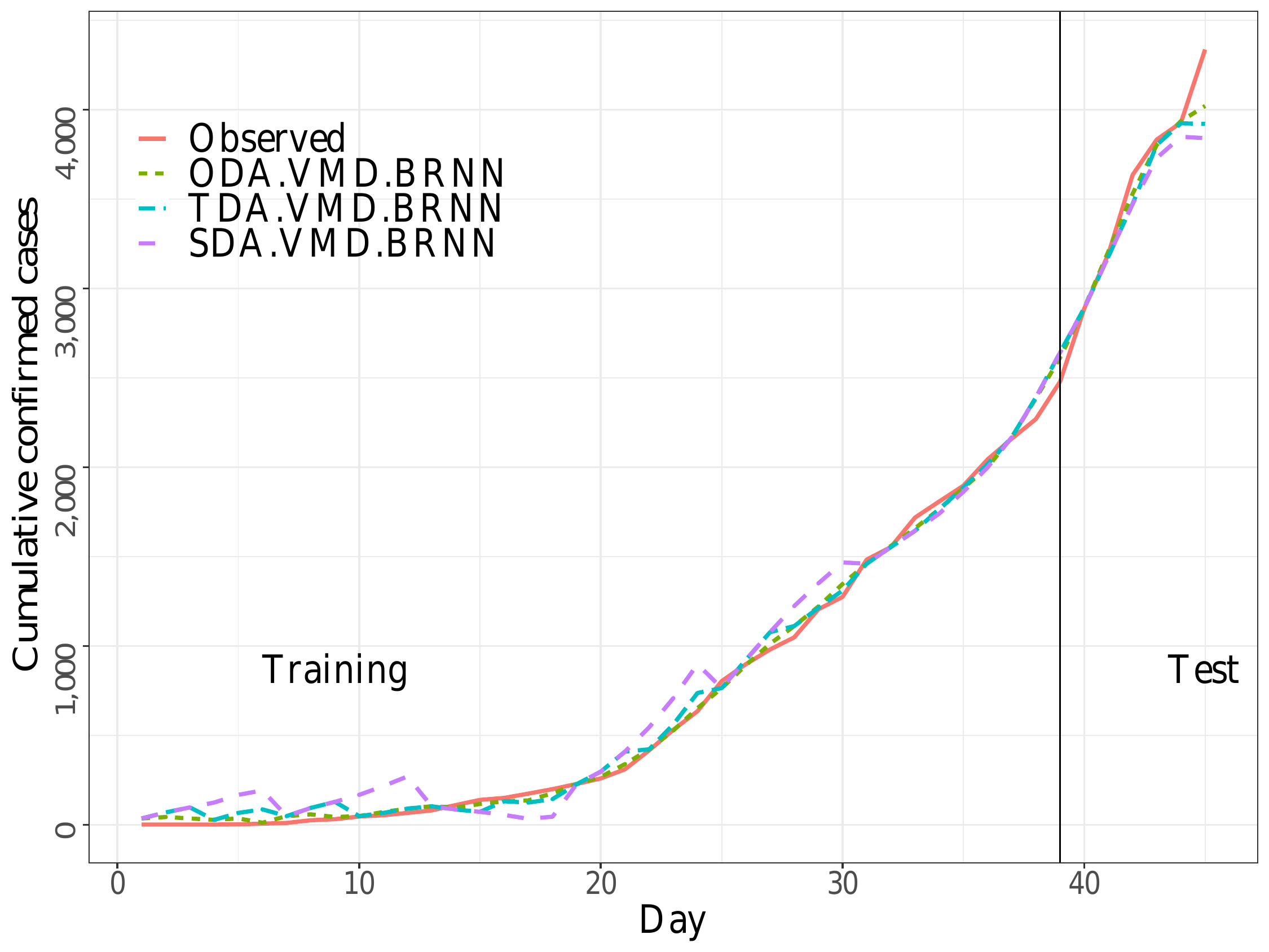}
    }
    \subfloat[CE \label{subfig:CE}]{
    \includegraphics[width=0.43\linewidth]{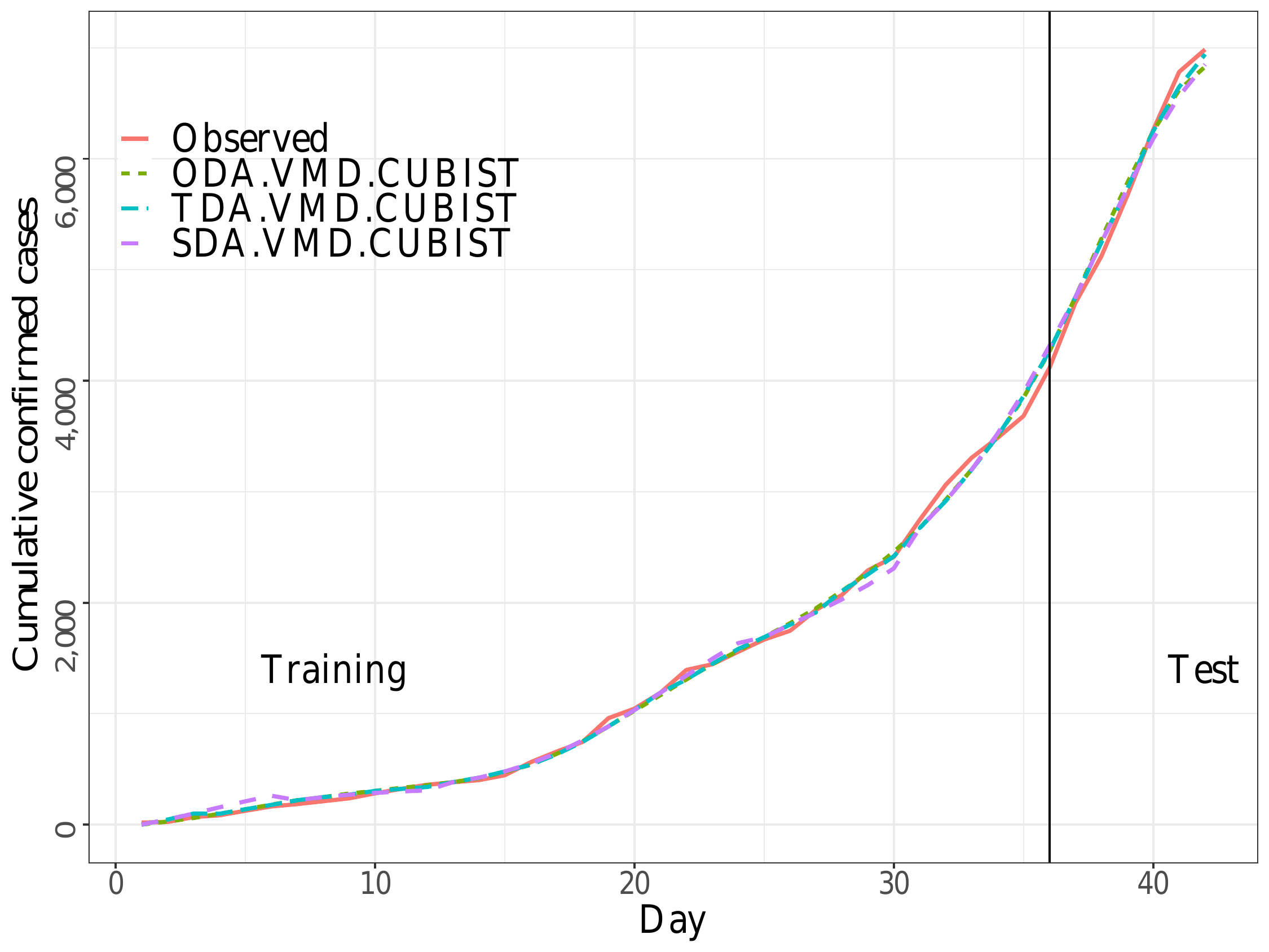}
    }
    
    \subfloat[PE \label{subfig:PE}]{
    \includegraphics[width=0.43\linewidth]{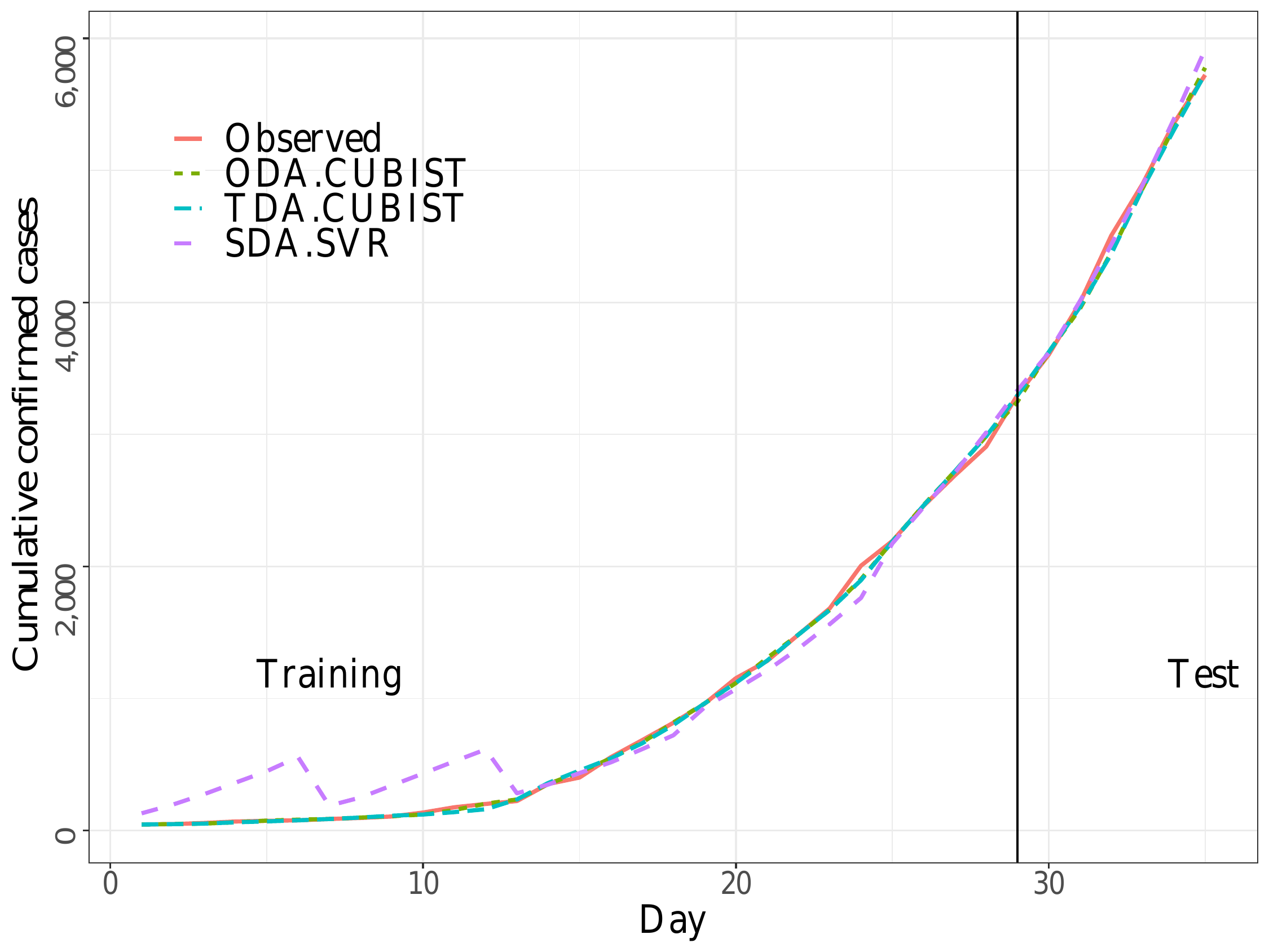}
    }
    \subfloat[RJ \label{subfig:RJ}]{
    \includegraphics[width=0.43\linewidth]{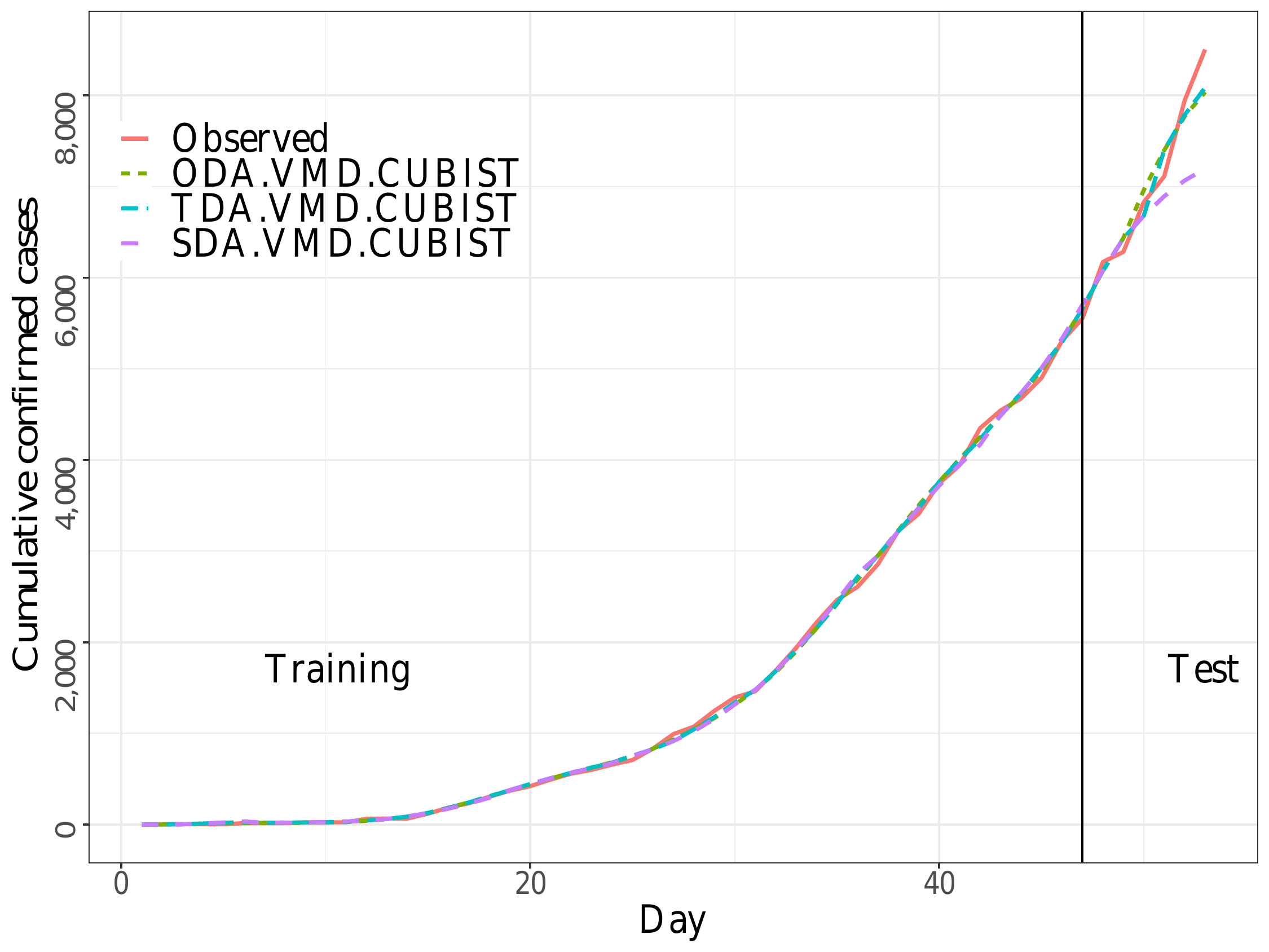}
    }
    
    \subfloat[SP \label{subfig:SP}]{
    \includegraphics[width=0.43\linewidth]{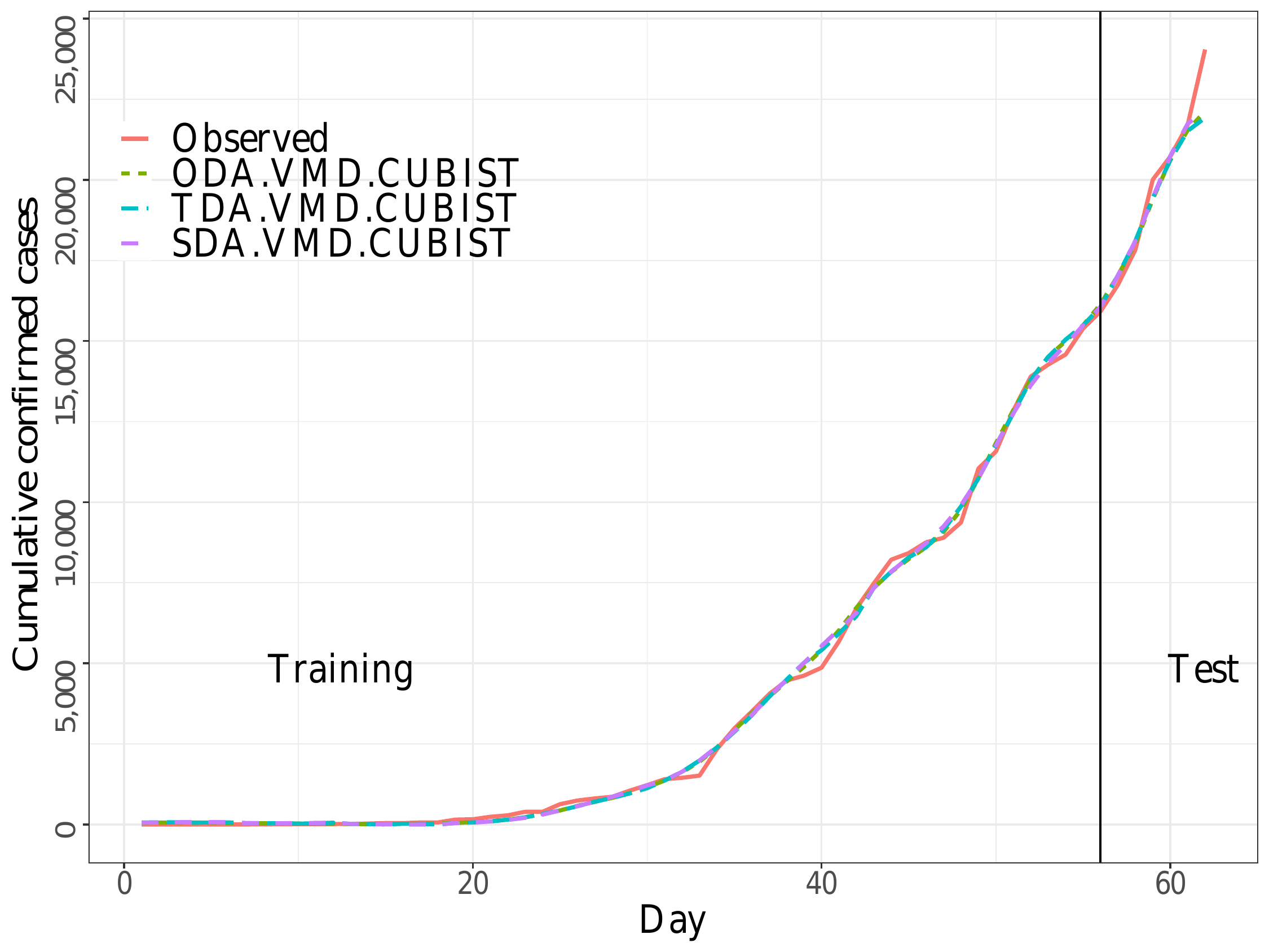}
    }
    \caption{Prediction versus observed COVID-19 cases for Brazilian States}
    \label{fig:PObrazil}
\end{figure}

\begin{figure}[htb!]
    \centering
    \subfloat[CA \label{subfig:CA}]{
    \includegraphics[width=0.43\linewidth]{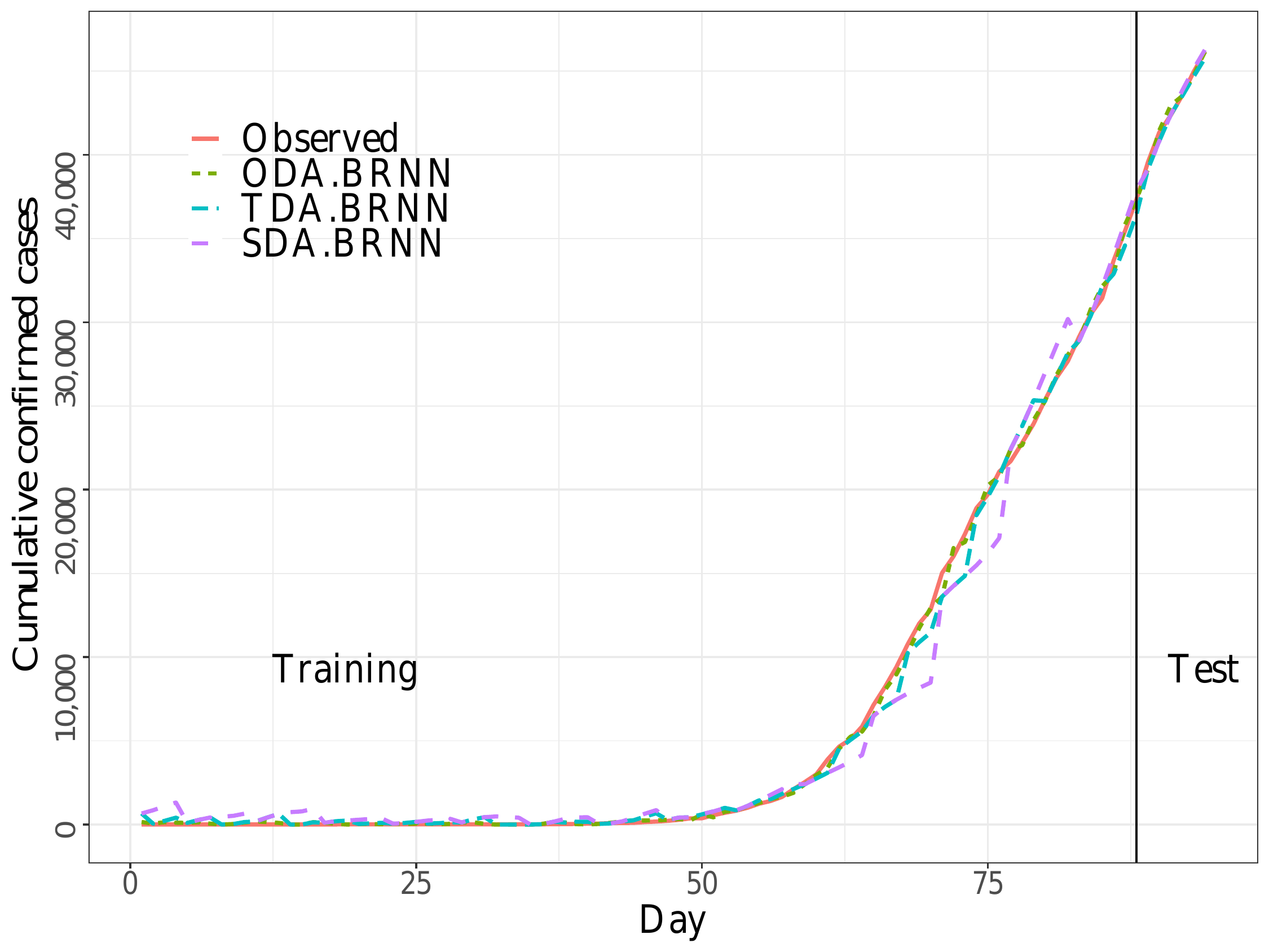}
    }
    \subfloat[IL \label{subfig:IL}]{
    \includegraphics[width=0.43\linewidth]{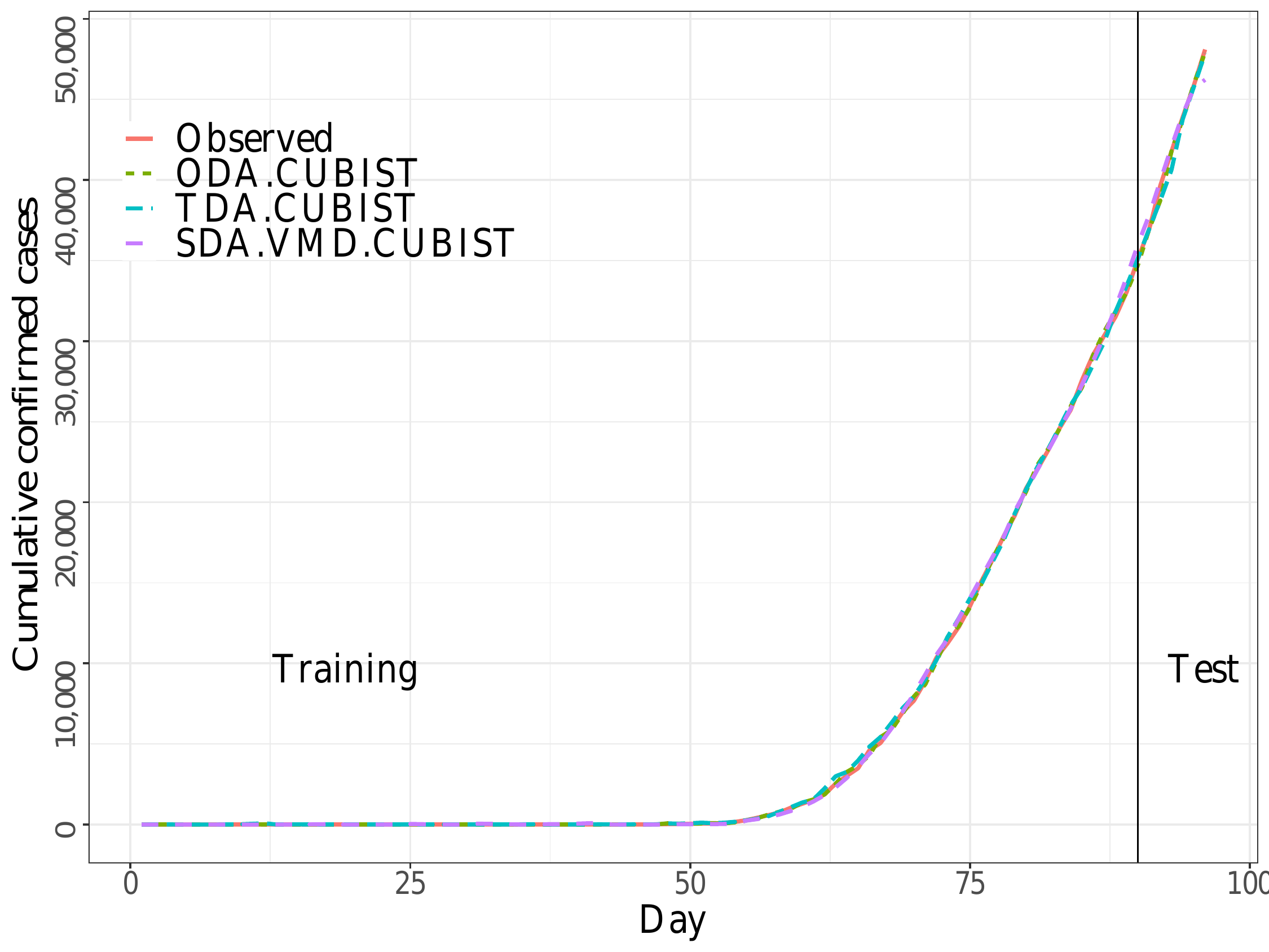}
    }

    \subfloat[MA \label{subfig:MA}]{
    \includegraphics[width=0.43\linewidth]{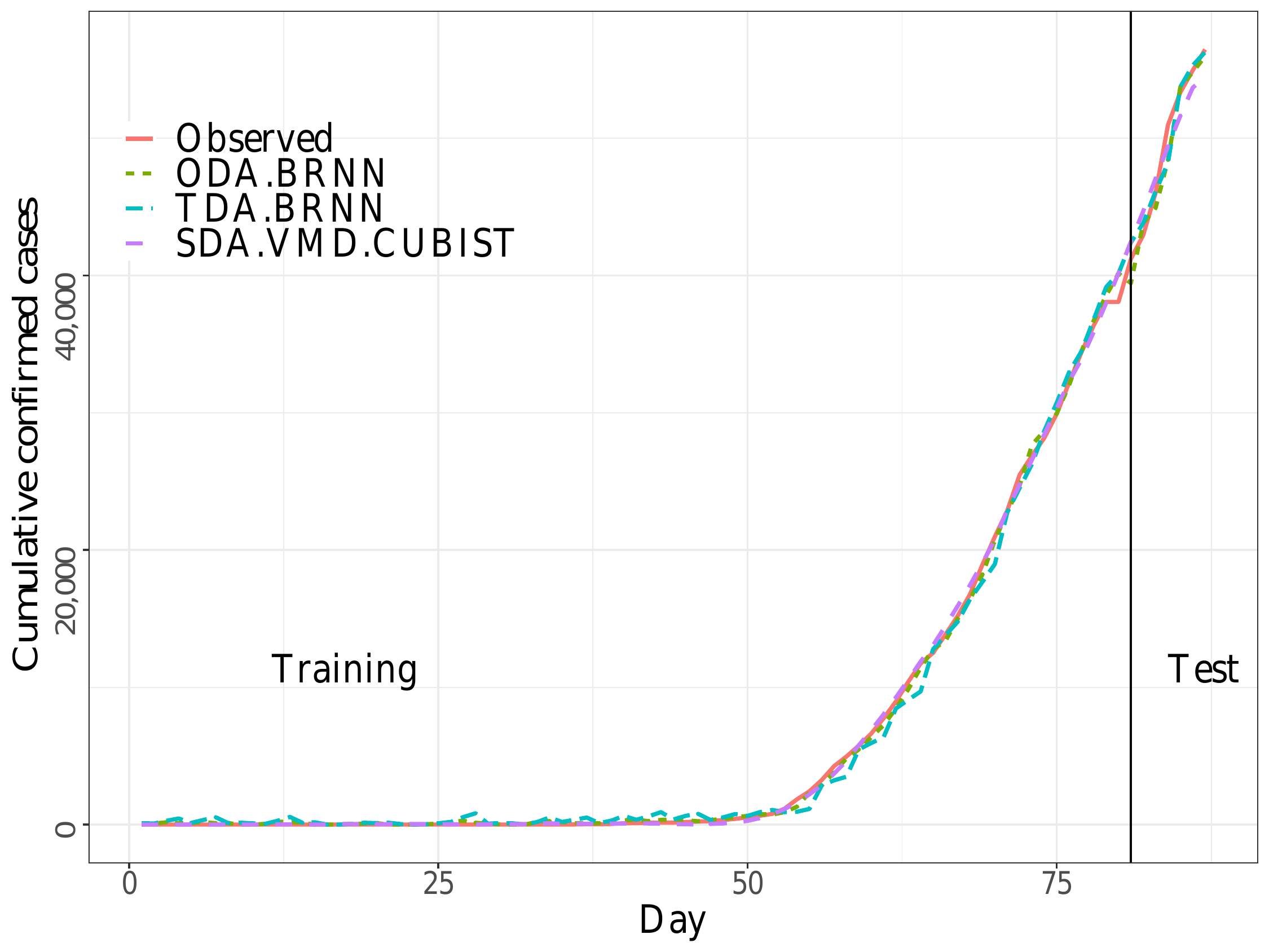}
    }
    \subfloat[NJ \label{subfig:NJ}]{
    \includegraphics[width=0.43\linewidth]{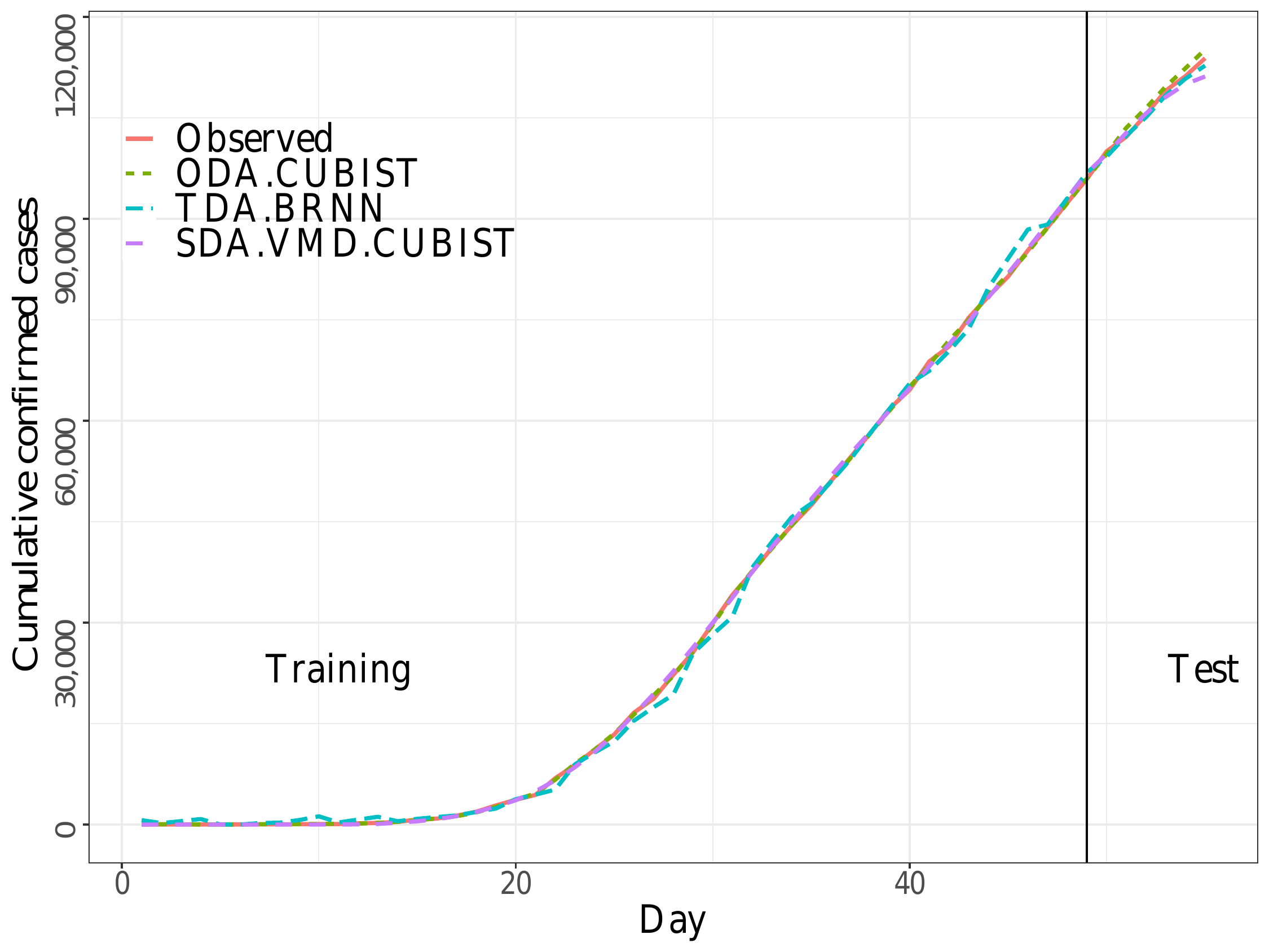}
    }
    
    \subfloat[NY \label{subfig:NY}]{
    \includegraphics[width=0.43\linewidth]{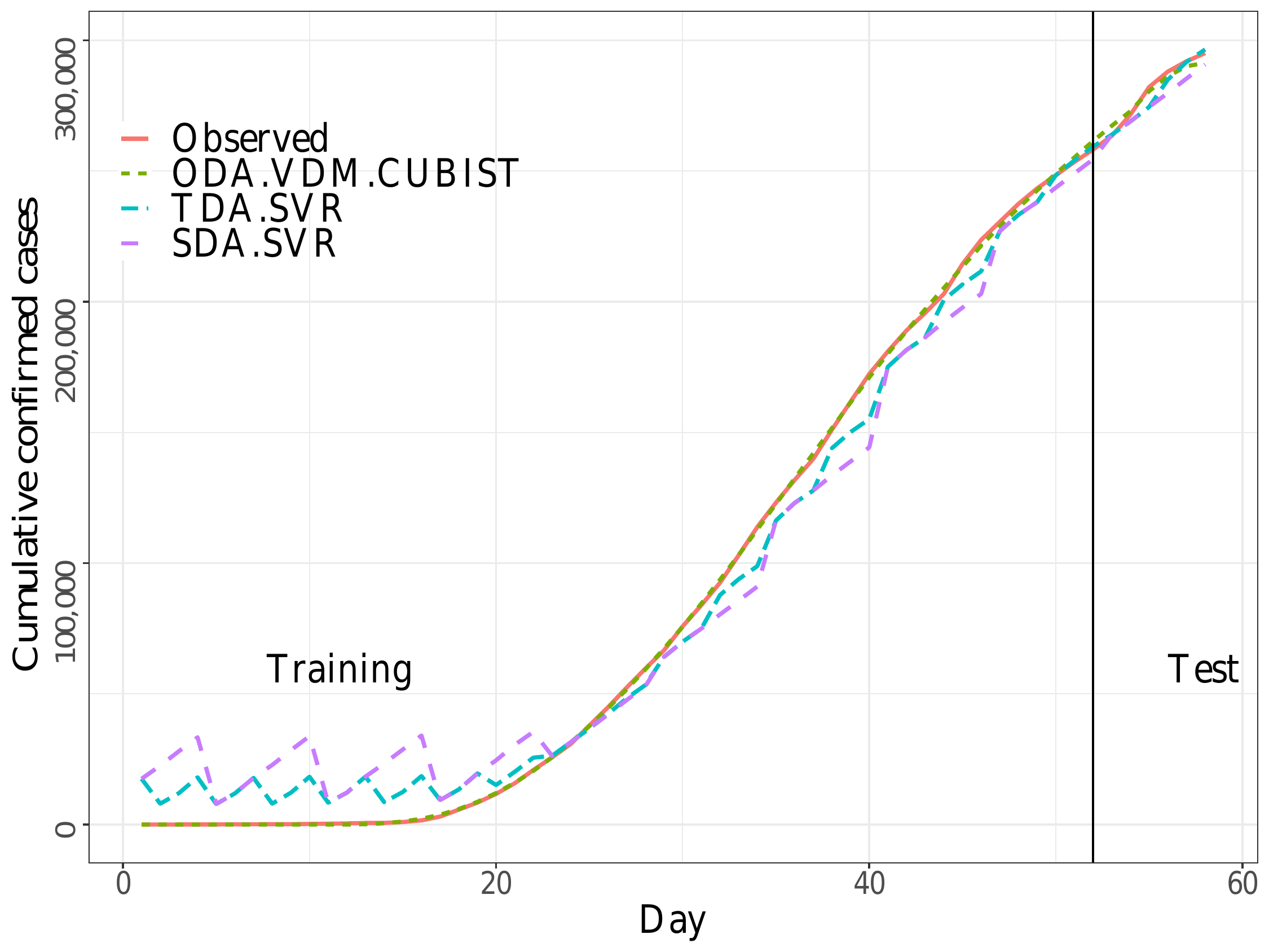}
    }
    \caption{Prediction versus observed COVID-19 cases for American States}
    \label{fig:POusa}
\end{figure}

According to the information depicted in Figures \ref{fig:PObrazil} and \ref{fig:POusa} it is possible to identify that the behavior of the data is learned by the evaluated models, which can forecasting compatible cases with the observed values. In most states, the good performance presented in the training stage persists in the test phase. In Figures \ref{subfig:AM}, \ref{subfig:PE}, \ref{subfig:CA}, and \ref{subfig:NY} the models presented some difficulties to capture the behavior of the data in the training stage, however in test phase the models could perform accurately presenting low errors. 

% BRA states

% USA states

\begin{figure}[htb!]
    \centering
    \includegraphics[width=\linewidth]{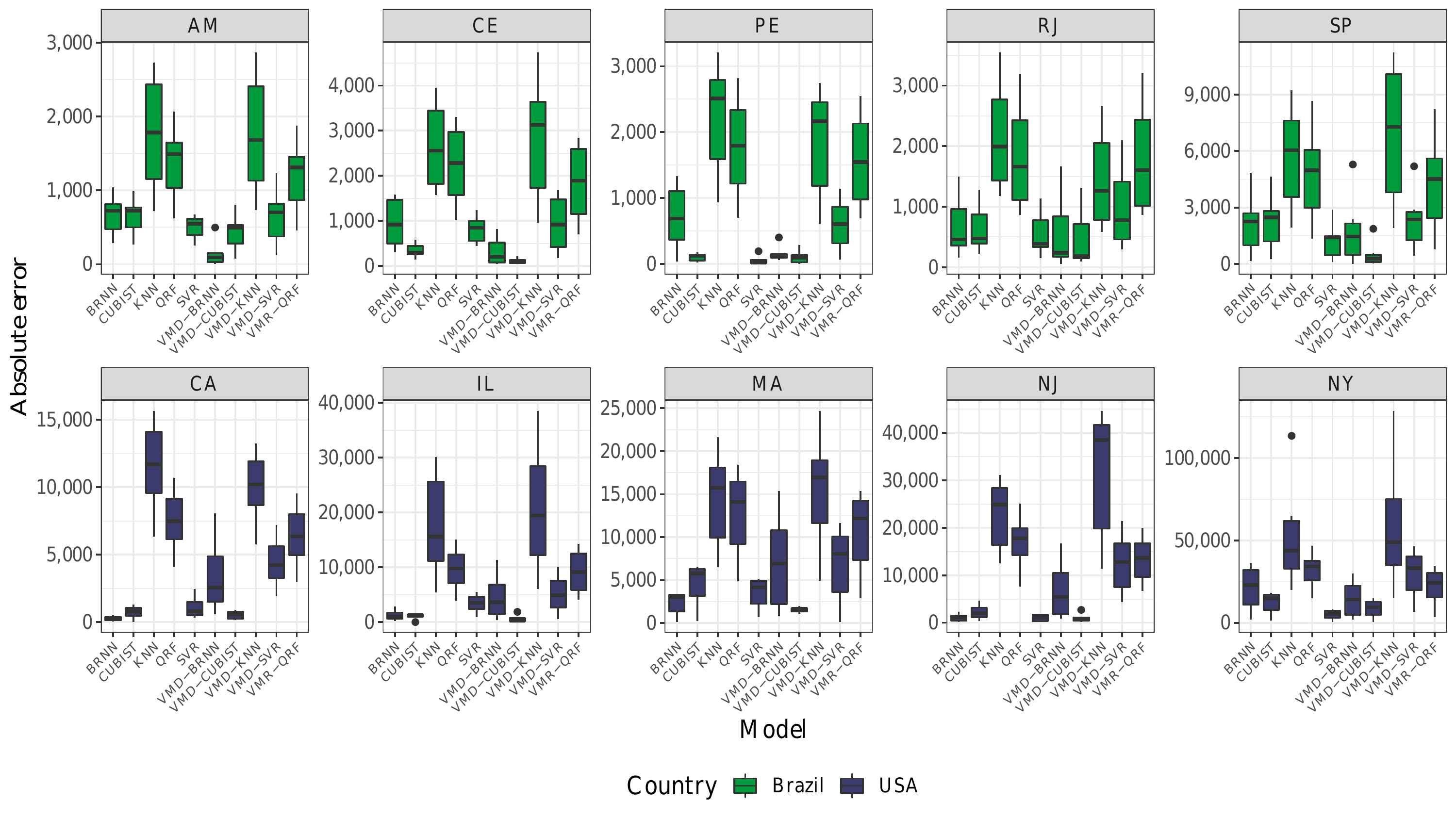}
    \caption{Box-plot for absolute error according to model and state for COVID-19 forecasting for SDA}
    \label{fig:error}
\end{figure}

Furthermore, Figure \ref{fig:error} presents the box-plots of test set forecasting errors in the SDA horizon for each model and each state. Due to the recursive strategy adopted, the SDA horizon was chosen to the analysis, once the errors tend to grow as the forecast horizon increases. The box diagram depicts the variation of absolute errors for each model, which reflects the stability of each model. In this context, the dots out of boxes are considered outliers errors.

% Box plot

Analyzing the box-plot, models with lower variation in the errors are indicated by the boxes with a smaller size. Figure \ref{fig:error} \comment{corroborates} the results presented in Tables \ref{tab:performancemeasure} and \ref{tab:performancemeasure2}. Models with lower errors achieve better stability, which means that the most appropriate model for each state can maintain a learning pattern, obtaining homogeneous forecasting errors.

\begin{figure}[htb!]
    \centering
    \includegraphics[width=0.8\linewidth]{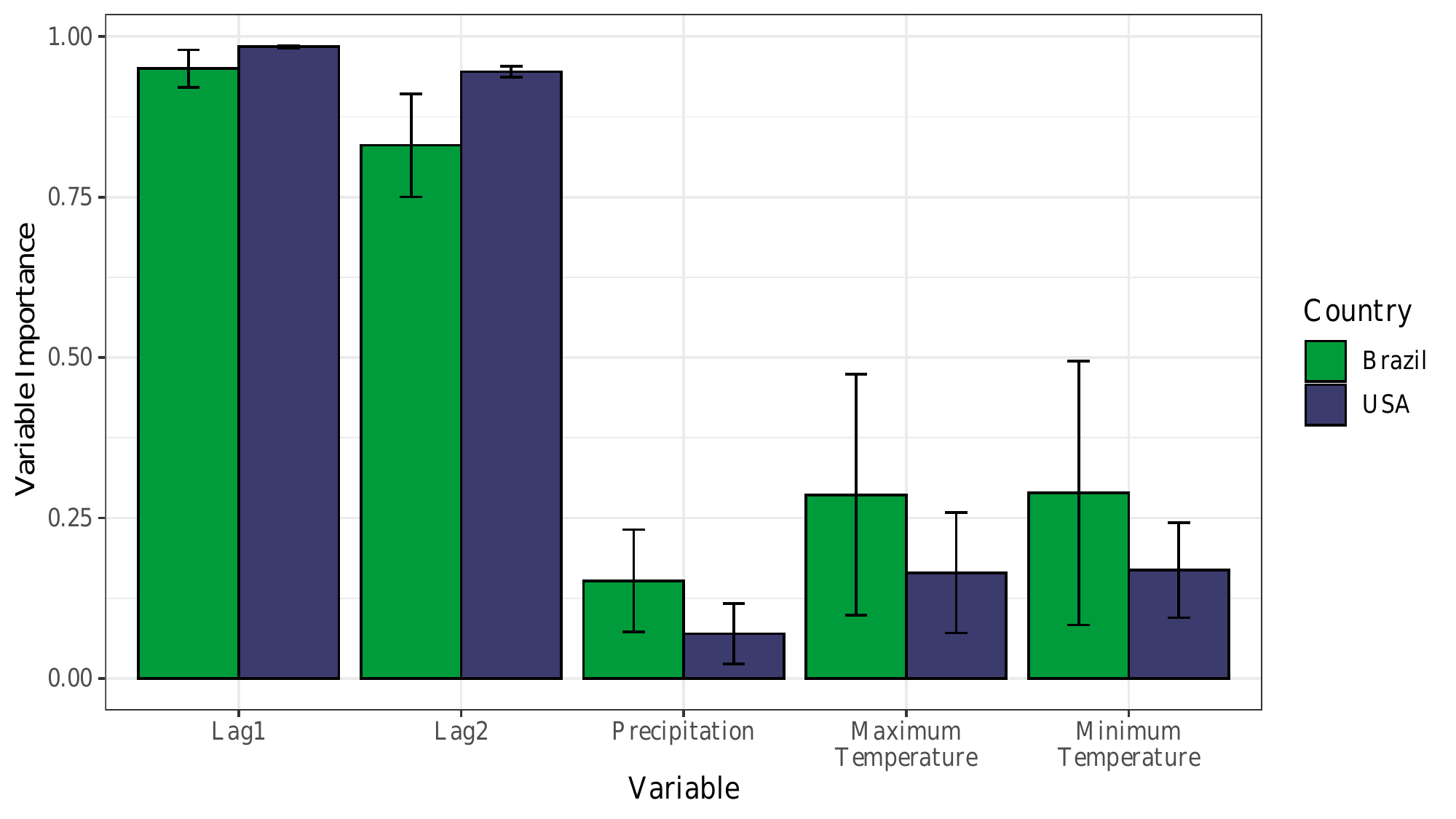}
    \caption{Variable importance for Brazil and USA}
    \label{fig:importance}
\end{figure}

\comment{The variable importance is an overall quantification of the relationship between the predictor variables (inputs) and the predicted value.} Finally, Figure \ref{fig:importance} is presented the variable importance of each input used to fit and train the models. As expected, the lag inputs present high importance due to their high correlation to the output. However, it is important to notice that climate data indeed presented some influence in predicting COVID-19 cumulative cases, especially in the Brazilian context, that the variance of the Temperature data reaches up to 50\% of importance. In other words, the climatic exogenous inputs are in some level relevant to the prediction of cumulative cases of COVID-19 in both Brazil's and USA's context \comment{for the five evaluated states.}

% Importance

    \section{Conclusion and Future Research \label{CONC}}

In this paper, machine learning approaches named BRNN, CUBIST, KNN, QRF, and SVR, as well as VMD approach, were employed in the task of forecasting one, three, and six-days-ahead the COVID-19 cumulative confirmed cases in five Brazilian states and five American states with a high daily incidence. The COVID-19 cumulative confirmed cases for AM, CE, PE, RJ, and SP states, as well as CA, IL, MA, NJ, and NY were used. The IP, sMAPE and RRMSE criteria were adopted to evaluate the performance of the compared approaches. The stability of out-of-sample errors was evaluated through box-plots. Further, the variable importance of the lag and climatic exogenous inputs were analyzed.

In respect of obtained results, it is possible to infer that CUBIST coupled with \comment{the} VMD model are suitable tools to forecast COVID-19 cases for most of the adopted states, once that these approaches were able to learn the non-linearities inherent to the evaluated epidemiological time series. Also, BRNN and SVR models deserve attention for the development of this task as well. Therefore, the ranking of models in all scenarios for Brazilian states is VMD--CUBIST, VMD--BRNN, SVR, CUBIST, VMD--SVR, BRNN, VMD--QRF, QRF, VMD--KNN, and KNN, and for USA states is VMD--CUBIST, BRNN, CUBIST, SVR, VMD--BRNN, VMD--SVR, VMD--QRF, QRF, KNN, and VMD--KNN. Also, looking for COVID-19 forecasts six-days-ahead, hybrid models are more suitable tools than non-decomposed models. Further, it was observed that climatic variables, such as temperature and precipitation indeed influence increasing the accuracy when predicting COVID-19 cases, wherein some cases climate inputs reached up to 50\% of importance in the forecasting model.

For future works, it is intended to adopt (i) deep learning approaches, (ii) different decomposition approaches, (iii) multi-objective optimization to tune hyperparameters of \comment{forecasting} models, and (iv) more climatic data and demographic features.

    \section*{CRediT Author Statement}

\textbf{Ramon Gomes da Silva:} Conceptualization, Methodology, Formal analysis, Validation, Writing - Original Draft, Writing - Review \& Editing.
\textbf{Matheus Henrique Dal Molin Ribeiro:} Conceptualization, Methodology, Formal analysis, Validation, Writing - Original Draft, Writing - Review \& Editing.
\textbf{Viviana Cocco Mariani:} Conceptualization, Writing - Review \& Editing.
\textbf{Leandro dos Santos Coelho:} Conceptualization, Writing - Review \& Editing.
    \section*{Declaration of Competing Interest}

The authors declare that they have no known competing financial interests or personal relationships that could have appeared to influence the work reported in this paper.
    \section*{Acknowledgments}

The authors would like to thank the National Council of Scientific and Technologic Development of Brazil -- CNPq (Grants number: 307958/2019-1-PQ, 307966/2019-4-PQ, 404659/2016-0-Univ, 405101/2016-3-Univ),  PRONEX `\textit{Funda\c{c}\~ao Arauc\'aria}' 042/2018, and \textit{Coordena\c{c}\~ao de Aperfei\c{c}oamento de Pessoal de N\'ivel Superior - Brasil} (CAPES) - Finance Code 001 for financial support of this work. Furthermore, the authors wish to thank the Editor and anonymous reviewers for their constructive comments and recommendations, which have significantly improved the presentation of this paper.
    
    % Referências
    \bibliographystyle{Misc/model3-num-names.bst}
    \bibliography{Misc/theBiblio}
    
    % Apêndices
    \appendix
    \newpage
    \section{Performance Measures \label{apendixA}}

Tables \ref{tab:performancemeasure} and \ref{tab:performancemeasure2} present the performance measures for each model in each state and forecasting horizon.

\setcounter{table}{0}

\begin{table}[htb!]
\centering
\caption{Performance measures for each evaluated model for Brazilian states}
\label{tab:performancemeasure}
\resizebox{\textwidth}{!}{%
\begin{tabular}{cccc|ccccc|ccccc}
\hline
\multirow{2}{*}{Country} & \multirow{2}{*}{State} & \multirow{2}{*}{\begin{tabular}[c]{@{}c@{}}Forecasting\\ Horizon\end{tabular}} & \multirow{2}{*}{Criteria} & \multicolumn{10}{c}{Model} \\ \cline{5-14}
 &  &  &  & BRNN & CUBIST & KNN & QRF & SVR & VMD--BRNN & VMD--CUBIST & VMD--KNN & VMD--QRF & VMD--SVR \\ \hline
\multirow{30}{*}{BRA} & \multirow{6}{*}{AM} & \multirow{2}{*}{ODA} & sMAPE & 11.59\% & 5.34\% & 50.92\% & 45.99\% & 5.40\% & \textbf{2.00\%} & 3.31\% & 38.17\% & 20.32\% & 4.30\% \\
 &  &  & RRMSE & 12.96\% & 5.88\% & 73.92\% & 65.17\% & 5.59\% & \textbf{3.83\%} & 4.25\% & 50.10\% & 23.55\% & 5.11\% \\
 &  & \multirow{2}{*}{TDA} & sMAPE & 14.54\% & 8.86\% & 52.37\% & 45.99\% & 7.65\% & \textbf{2.68\%} & 6.09\% & 51.73\% & 24.93\% & 9.58\% \\
 &  &  & RRMSE & 17.00\% & 11.05\% & 77.14\% & 65.17\% & 9.20\% & \textbf{5.18\%} & 7.77\% & 75.87\% & 30.25\% & 11.99\% \\
 &  & \multirow{2}{*}{SDA} & sMAPE & 19.62\% & 19.15\% & 63.80\% & 45.99\% & 14.48\% & \textbf{3.67\%} & 11.26\% & 63.53\% & 38.35\% & 18.68\% \\
 &  &  & RRMSE & 24.11\% & 23.31\% & 102.85\% & 65.17\% & 16.62\% & \textbf{6.29\%} & 14.50\% & 102.53\% & 52.68\% & 24.83\% \\ \cline{2-14}
 & \multirow{6}{*}{CE} & \multirow{2}{*}{ODA} & sMAPE & 7.97\% & 2.70\% & 53.33\% & 45.42\% & 5.49\% & 2.55\% & \textbf{1.83\%} & 40.75\% & 17.82\% & 3.26\% \\
 &  &  & RRMSE & 9.48\% & 3.89\% & 78.05\% & 64.86\% & 5.52\% & 3.52\% & \textbf{2.09\%} & 56.19\% & 20.56\% & 3.85\% \\
 &  & \multirow{2}{*}{TDA} & sMAPE & 13.17\% & 4.24\% & 55.23\% & 45.42\% & 8.46\% & 3.46\% & \textbf{1.23\%} & 52.98\% & 22.64\% & 8.03\% \\
 &  &  & RRMSE & 16.46\% & 4.26\% & 82.59\% & 64.86\% & 8.85\% & 5.35\% & \textbf{1.41\%} & 79.73\% & 26.70\% & 9.57\% \\
 &  & \multirow{2}{*}{SDA} & sMAPE & 16.69\% & 5.91\% & 56.75\% & 45.42\% & 14.36\% & 5.08\% & \textbf{1.82\%} & 62.54\% & 35.71\% & 16.20\% \\
 &  &  & RRMSE & 21.78\% & 6.62\% & 86.22\% & 64.86\% & 16.85\% & 7.71\% & \textbf{2.08\%} & 101.64\% & 49.40\% & 22.04\% \\ \cline{2-14}
 & \multirow{6}{*}{PE} & \multirow{2}{*}{ODA} & sMAPE & 10.43\% & \textbf{1.14\%} & 54.97\% & 45.27\% & 1.22\% & 2.54\% & 2.09\% & 37.56\% & 26.69\% & 1.96\% \\
 &  &  & RRMSE & 13.54\% & \textbf{1.36\%} & 83.80\% & 66.03\% & 1.86\% & 3.32\% & 2.39\% & 51.61\% & 32.54\% & 2.65\% \\
 &  & \multirow{2}{*}{TDA} & sMAPE & 12.21\% & \textbf{1.06\%} & 60.19\% & 45.27\% & 1.42\% & 3.30\% & 1.78\% & 44.05\% & 30.83\% & 6.11\% \\
 &  &  & RRMSE & 16.57\% & \textbf{1.39\%} & 94.20\% & 66.03\% & 2.33\% & 4.37\% & 2.11\% & 62.59\% & 40.21\% & 7.94\% \\
 &  & \multirow{2}{*}{SDA} & sMAPE & 15.32\% & 2.08\% & 61.09\% & 45.27\% & \textbf{1.05\%} & 3.28\% & 2.16\% & 48.27\% & 39.07\% & 12.76\% \\
 &  &  & RRMSE & 21.34\% & 2.53\% & 96.22\% & 66.03\% & \textbf{1.81\%} & 4.15\% & 3.08\% & 71.19\% & 55.22\% & 17.24\% \\ \cline{2-14}
 & \multirow{6}{*}{RJ} & \multirow{2}{*}{ODA} & sMAPE & 4.87\% & 3.65\% & 34.70\% & 28.70\% & 3.54\% & 3.61\% & \textbf{3.05\%} & 16.88\% & 15.87\% & 3.80\% \\
 &  &  & RRMSE & 5.82\% & 4.37\% & 46.38\% & 38.06\% & 4.62\% & 5.32\% & \textbf{3.60\%} & 21.62\% & 19.64\% & 5.24\% \\
 &  & \multirow{2}{*}{TDA} & sMAPE & 6.68\% & 5.62\% & 34.95\% & 28.70\% & 5.34\% & 5.58\% & \textbf{2.73\%} & 18.69\% & 20.51\% & 7.67\% \\
 &  &  & RRMSE & 9.23\% & 7.00\% & 46.81\% & 38.06\% & 6.72\% & 9.02\% & \textbf{3.30\%} & 23.93\% & 25.51\% & 9.88\% \\
 &  & \multirow{2}{*}{SDA} & sMAPE & 9.38\% & 8.95\% & 34.95\% & 28.70\% & 7.57\% & 7.67\% & \textbf{3.02\%} & 21.92\% & 28.10\% & 14.24\% \\
 &  &  & RRMSE & 12.73\% & 11.40\% & 46.81\% & 38.06\% & 9.83\% & 12.35\% & \textbf{3.95\%} & 28.99\% & 37.48\% & 19.27\% \\ \cline{2-14}
 & \multirow{6}{*}{SP} & \multirow{2}{*}{ODA} & sMAPE & 4.70\% & 3.58\% & 29.85\% & 26.12\% & 2.81\% & 3.73\% & \textbf{2.57\%} & 32.58\% & 20.39\% & 3.91\% \\
 &  &  & RRMSE & 6.81\% & 4.82\% & 39.71\% & 34.83\% & 4.22\% & 5.85\% & \textbf{4.22\%} & 43.82\% & 27.35\% & 5.06\% \\
 &  & \multirow{2}{*}{TDA} & sMAPE & 6.44\% & 4.56\% & 29.85\% & 26.12\% & 3.17\% & 6.14\% & \textbf{2.79\%} & 37.21\% & 21.95\% & 6.65\% \\
 &  &  & RRMSE & 9.62\% & 6.45\% & 39.71\% & 34.83\% & 4.74\% & 10.04\% & \textbf{4.62\%} & 52.39\% & 29.47\% & 8.73\% \\
 &  & \multirow{2}{*}{SDA} & sMAPE & 10.67\% & 11.29\% & 32.05\% & 26.12\% & 5.95\% & 8.71\% & \textbf{2.42\%} & 40.67\% & 23.02\% & 11.82\% \\
 &  &  & RRMSE & 14.65\% & 14.92\% & 43.16\% & 34.83\% & 8.19\% & 13.68\% & \textbf{4.10\%} & 58.44\% & 31.18\% & 15.78\% \\ \hline
\end{tabular}%
}
\end{table}

\begin{table}[htb!]
\centering
\caption{Performance measures for each evaluated model for American states}
\label{tab:performancemeasure2}
\resizebox{\textwidth}{!}{%
\begin{tabular}{cccc|ccccc|ccccc}
\hline
\multirow{2}{*}{Country} & \multirow{2}{*}{State} & \multirow{2}{*}{\begin{tabular}[c]{@{}c@{}}Forecasting\\ Horizon\end{tabular}} & \multirow{2}{*}{Criteria} & \multicolumn{10}{c}{Model} \\ \cline{5-14}
 &  &  &  & BRNN & CUBIST & KNN & QRF & SVR & VMD--BRNN & VMD--CUBIST & VMD--KNN & VMD--QRF & VMD--SVR \\ \hline
\multirow{30}{*}{USA} & \multirow{6}{*}{CA} & \multirow{2}{*}{ODA} & sMAPE & \textbf{0.56\%} & 0.90\% & 26.37\% & 19.05\% & 1.80\% & 2.19\% & 0.80\% & 24.23\% & 11.44\% & 3.60\% \\
 &  &  & RRMSE & \textbf{0.73\%} & 0.96\% & 31.31\% & 22.12\% & 1.96\% & 2.63\% & 1.07\% & 28.42\% & 12.87\% & 3.76\% \\
 &  & \multirow{2}{*}{TDA} & sMAPE & \textbf{0.66\%} & 1.10\% & 28.87\% & 19.05\% & 2.40\% & 4.66\% & 0.69\% & 25.63\% & 13.07\% & 6.11\% \\
 &  &  & RRMSE & \textbf{0.78\%} & 1.22\% & 35.11\% & 22.12\% & 2.97\% & 6.08\% & 0.84\% & 30.44\% & 14.62\% & 6.74\% \\
 &  & \multirow{2}{*}{SDA} & sMAPE & \textbf{0.62\%} & 1.87\% & 30.85\% & 19.05\% & 2.41\% & 8.20\% & 1.21\% & 26.23\% & 15.89\% & 10.75\% \\
 &  &  & RRMSE & \textbf{0.72\%} & 2.18\% & 38.04\% & 22.12\% & 3.01\% & 10.86\% & 1.40\% & 31.31\% & 18.44\% & 12.38\% \\ \cline{2-14}
 & \multirow{6}{*}{IL} & \multirow{2}{*}{ODA} & sMAPE & 1.07\% & \textbf{0.54\%} & 34.26\% & 25.04\% & 3.05\% & 3.16\% & 1.83\% & 43.71\% & 17.74\% & 3.80\% \\
 &  &  & RRMSE & 1.24\% & \textbf{0.92\%} & 44.21\% & 31.31\% & 3.13\% & 4.43\% & 2.63\% & 60.63\% & 21.50\% & 4.49\% \\
 &  & \multirow{2}{*}{TDA} & sMAPE & 1.84\% & \textbf{1.05\%} & 36.63\% & 25.04\% & 4.73\% & 6.33\% & 1.55\% & 49.22\% & 19.41\% & 6.98\% \\
 &  &  & RRMSE & 2.39\% & \textbf{1.51\%} & 47.86\% & 31.31\% & 5.18\% & 8.95\% & 2.07\% & 70.55\% & 23.08\% & 8.53\% \\
 &  & \multirow{2}{*}{SDA} & sMAPE & 2.89\% & 2.58\% & 52.64\% & 25.04\% & 8.10\% & 11.03\% & \textbf{1.42\%} & 67.54\% & 23.61\% & 12.43\% \\
 &  &  & RRMSE & 3.78\% & 3.00\% & 78.46\% & 31.31\% & 9.53\% & 15.66\% & \textbf{2.04\%} & 107.77\% & 29.59\% & 16.27\% \\ \cline{2-14}
 & \multirow{6}{*}{MA} & \multirow{2}{*}{ODA} & sMAPE & \textbf{1.90\%} & 2.45\% & 30.79\% & 28.07\% & 3.01\% & 4.51\% & 2.65\% & 26.67\% & 17.49\% & 3.91\% \\
 &  &  & RRMSE & \textbf{2.42\%} & 3.46\% & 39.30\% & 35.70\% & 3.59\% & 5.39\% & 2.69\% & 33.79\% & 21.69\% & 4.77\% \\
 &  & \multirow{2}{*}{TDA} & sMAPE & \textbf{1.59\%} & 3.41\% & 31.52\% & 28.07\% & 3.35\% & 8.25\% & 2.34\% & 31.47\% & 19.83\% & 7.12\% \\
 &  &  & RRMSE & \textbf{2.39\%} & 5.34\% & 40.65\% & 35.70\% & 4.58\% & 10.86\% & 2.41\% & 40.87\% & 24.21\% & 8.82\% \\
 &  & \multirow{2}{*}{SDA} & sMAPE & 4.38\% & 8.98\% & 32.54\% & 28.07\% & 6.92\% & 14.65\% & \textbf{3.08\%} & 35.52\% & 22.66\% & 13.85\% \\
 &  &  & RRMSE & 5.33\% & 10.98\% & 42.30\% & 35.70\% & 8.18\% & 20.17\% & \textbf{3.16\%} & 47.50\% & 28.56\% & 18.13\% \\ \cline{2-14}
 & \multirow{6}{*}{NJ} & \multirow{2}{*}{ODA} & sMAPE & 0.97\% & \textbf{0.88\%} & 21.41\% & 17.87\% & 0.94\% & 1.63\% & 0.99\% & 25.40\% & 9.88\% & 3.78\% \\
 &  &  & RRMSE & 1.01\% & \textbf{0.93\%} & 24.93\% & 20.38\% & 1.03\% & 2.18\% & 1.32\% & 30.87\% & 10.99\% & 3.92\% \\
 &  & \multirow{2}{*}{TDA} & sMAPE & \textbf{0.55\%} & 1.25\% & 22.03\% & 18.20\% & 1.09\% & 3.52\% & 0.83\% & 28.62\% & 11.17\% & 6.75\% \\
 &  &  & RRMSE & \textbf{0.62\%} & 1.48\% & 25.81\% & 20.82\% & 1.17\% & 4.74\% & 1.21\% & 35.63\% & 12.36\% & 7.47\% \\
 &  & \multirow{2}{*}{SDA} & sMAPE & 1.02\% & 1.99\% & 23.76\% & 18.20\% & 0.94\% & 6.54\% & \textbf{0.91\%} & 35.12\% & 13.20\% & 12.35\% \\
 &  &  & RRMSE & 1.27\% & 2.40\% & 28.32\% & 20.82\% & 1.14\% & 8.95\% & \textbf{1.23\%} & 45.47\% & 15.09\% & 14.70\% \\ \cline{2-14}
 & \multirow{6}{*}{NY} & \multirow{2}{*}{ODA} & sMAPE & 3.26\% & 1.26\% & 16.81\% & 12.09\% & 1.02\% & 1.46\% & \textbf{0.84\%} & 18.44\% & 5.51\% & 3.27\% \\
 &  &  & RRMSE & 3.63\% & 1.43\% & 19.88\% & 13.64\% & 1.23\% & 1.64\% & \textbf{0.92\%} & 23.44\% & 6.30\% & 3.38\% \\
 &  & \multirow{2}{*}{TDA} & sMAPE & 5.40\% & 2.20\% & 18.23\% & 12.09\% & \textbf{0.92\%} & 2.76\% & 1.10\% & 21.34\% & 6.39\% & 5.85\% \\
 &  &  & RRMSE & 6.43\% & 2.69\% & 22.36\% & 13.64\% & \textbf{1.27\%} & 3.31\% & 1.38\% & 27.91\% & 7.28\% & 6.62\% \\
 &  & \multirow{2}{*}{SDA} & sMAPE & 7.64\% & 4.54\% & 21.02\% & 12.09\% & \textbf{1.75\%} & 5.23\% & 3.05\% & 24.06\% & 7.96\% & 10.97\% \\
 &  &  & RRMSE & 9.37\% & 5.30\% & 26.69\% & 13.64\% & \textbf{2.05\%} & 6.65\% & 3.65\% & 31.38\% & 9.32\% & 12.93\% \\ \hline
\end{tabular}%
}
\end{table}

    \section{Hyperparameters \label{apendixB}}

Tables \ref{tab:hyper} and \ref{tab:hyper2} present the hyperparameters obtained by grid-search for the models employed in this paper.

\setcounter{table}{0}

\begin{table}[htb!]
% \scriptsize
\centering
\caption{Hyperparameters selected by grid-search for each evaluated model for Brazilian states}
\label{tab:hyper}
\resizebox{\textwidth}{!}{%
\begin{tabular}{cclc|cc|c|c|c}
\hline
\multirow{2}{*}{Country} &
  \multirow{2}{*}{State} &
  \multicolumn{1}{c}{\multirow{2}{*}{Component}} &
  BRNN &
  \multicolumn{2}{c|}{CUBIST} &
  KNN &
  QRF &
  SVR \\ \cline{4-9}
 &
   &
   &
  \# of Neurons &
  \# of Committees &
  \# of Instances &
  \# of Neighbors &
  \begin{tabular}[c]{@{}c@{}}\begin{tabular}[c]{@{}c@{}}\# of Randomly\\Selected Predictors\end{tabular}\end{tabular} &
  Cost \\ \hline
\multirow{30}{*}{BRA} & \multirow{6}{*}{AM} & IMF$_1$        & 4 & 1  & 0 & 9  & 5 & 1 \\
                      &                     & IMF$_2$        & 5 & 20 & 5 & 5  & 5 & 1 \\
                      &                     & IMF$_3$        & 3 & 20 & 0 & 5  & 5 & 1 \\
                      &                     & IMF$_4$        & 5 & 10 & 0 & 7  & 5 & 1 \\
                      &                     & IMF$_5$        & 4 & 1  & 5 & 5  & 5 & 1 \\
                      &                     & Non-decomposed & 3 & 1  & 5 & 5  & 4 & 1 \\ \cline{2-9}
                      & \multirow{6}{*}{CE} & IMF$_1$        & 3 & 1  & 5 & 13 & 2 & 1 \\
                      &                     & IMF$_2$        & 5 & 20 & 5 & 5  & 5 & 1 \\
                      &                     & IMF$_3$        & 5 & 10 & 9 & 13 & 5 & 1 \\
                      &                     & IMF$_4$        & 5 & 10 & 9 & 5  & 5 & 1 \\
                      &                     & IMF$_5$        & 5 & 10 & 0 & 5  & 5 & 1 \\
                      &                     & Non-decomposed & 1 & 1  & 9 & 5  & 4 & 1 \\ \cline{2-9}
                      & \multirow{6}{*}{PE} & IMF$_1$        & 2 & 10 & 0 & 13 & 3 & 1 \\
                      &                     & IMF$_2$        & 5 & 20 & 5 & 5  & 3 & 1 \\
                      &                     & IMF$_3$        & 1 & 20 & 5 & 13 & 3 & 1 \\
                      &                     & IMF$_4$        & 5 & 1  & 9 & 5  & 3 & 1 \\
                      &                     & IMF$_5$        & 5 & 10 & 9 & 11 & 5 & 1 \\
                      &                     & Non-decomposed & 5 & 10 & 0 & 5  & 4 & 1 \\ \cline{2-9}
                      & \multirow{6}{*}{RJ} & IMF$_1$        & 3 & 10 & 0 & 5  & 5 & 1 \\
                      &                     & IMF$_2$        & 1 & 1  & 0 & 5  & 4 & 1 \\
                      &                     & IMF$_3$        & 4 & 20 & 5 & 5  & 5 & 1 \\
                      &                     & IMF$_4$        & 5 & 1  & 5 & 5  & 5 & 1 \\
                      &                     & IMF$_5$        & 5 & 1  & 5 & 5  & 5 & 1 \\
                      &                     & Non-decomposed & 1 & 20 & 5 & 5  & 4 & 1 \\ \cline{2-9}
                      & \multirow{6}{*}{SP} & IMF$_1$        & 2 & 10 & 0 & 11 & 4 & 1 \\
                      &                     & IMF$_2$        & 5 & 1  & 9 & 5  & 4 & 1 \\
                      &                     & IMF$_3$        & 5 & 10 & 5 & 13 & 4 & 1 \\
                      &                     & IMF$_4$        & 5 & 20 & 0 & 5  & 5 & 1 \\
                      &                     & IMF$_5$        & 1 & 10 & 5 & 5  & 5 & 1 \\
                      &                     & Non-decomposed & 1 & 10 & 0 & 5  & 5 & 1 \\ \hline
\end{tabular}%
}
\end{table}

\begin{table}[htb!]
% \scriptsize
\centering
\caption{Hyperparameters selected by grid-search for each evaluated model for American states}
\label{tab:hyper2}
\resizebox{\textwidth}{!}{%
\begin{tabular}{cclc|cc|c|c|c}
\hline
\multirow{2}{*}{Country} &
  \multirow{2}{*}{State} &
  \multicolumn{1}{c}{\multirow{2}{*}{Component}} &
  BRNN &
  \multicolumn{2}{c|}{CUBIST} &
  KNN &
  QRF &
  SVR \\ \cline{4-9}
 &
   &
   &
  \# of Neurons &
  \# of Committees &
  \# of Instances &
  \# of Neighbors &
  \begin{tabular}[c]{@{}c@{}}\begin{tabular}[c]{@{}c@{}}\# of Randomly\\Selected Predictors\end{tabular}\end{tabular} &
  Cost \\ \hline
\multirow{30}{*}{USA} & \multirow{6}{*}{CA} & IMF$_1$        & 1 & 1  & 9 & 5  & 4 & 1 \\
                      &                     & IMF$_2$        & 1 & 1  & 0 & 5  & 4 & 1 \\
                      &                     & IMF$_3$        & 4 & 1  & 9 & 11 & 3 & 1 \\
                      &                     & IMF$_4$        & 5 & 20 & 0 & 9  & 5 & 1 \\
                      &                     & IMF$_5$        & 5 & 1  & 5 & 5  & 5 & 1 \\
                      &                     & Non-decomposed & 1 & 20 & 5 & 5  & 4 & 1 \\ \cline{2-9}
                      & \multirow{6}{*}{IL} & IMF$_1$        & 5 & 20 & 5 & 5  & 5 & 1 \\
                      &                     & IMF$_2$        & 1 & 20 & 5 & 5  & 4 & 1 \\
                      &                     & IMF$_3$        & 5 & 20 & 5 & 5  & 3 & 1 \\
                      &                     & IMF$_4$        & 5 & 20 & 9 & 5  & 5 & 1 \\
                      &                     & IMF$_5$        & 5 & 10 & 0 & 5  & 5 & 1 \\
                      &                     & Non-decomposed & 1 & 20 & 5 & 5  & 4 & 1 \\ \cline{2-9}
                      & \multirow{6}{*}{MA} & IMF$_1$        & 3 & 1  & 5 & 5  & 5 & 1 \\
                      &                     & IMF$_2$        & 1 & 20 & 5 & 5  & 4 & 1 \\
                      &                     & IMF$_3$        & 4 & 20 & 5 & 5  & 4 & 1 \\
                      &                     & IMF$_4$        & 4 & 20 & 5 & 13 & 5 & 1 \\
                      &                     & IMF$_5$        & 5 & 1  & 0 & 5  & 5 & 1 \\
                      &                     & Non-decomposed & 1 & 20 & 5 & 5  & 5 & 1 \\ \cline{2-9}
                      & \multirow{6}{*}{NJ} & IMF$_1$        & 1 & 10 & 0 & 5  & 5 & 1 \\
                      &                     & IMF$_2$        & 5 & 20 & 5 & 5  & 4 & 1 \\
                      &                     & IMF$_3$        & 5 & 10 & 9 & 5  & 4 & 1 \\
                      &                     & IMF$_4$        & 4 & 10 & 9 & 13 & 5 & 1 \\
                      &                     & IMF$_5$        & 5 & 1  & 0 & 5  & 5 & 1 \\
                      &                     & Non-decomposed & 1 & 20 & 5 & 5  & 5 & 1 \\ \cline{2-9}
                      & \multirow{6}{*}{NY} & IMF$_1$        & 1 & 1  & 0 & 13 & 5 & 1 \\
                      &                     & IMF$_2$        & 5 & 20 & 5 & 5  & 5 & 1 \\
                      &                     & IMF$_3$        & 5 & 1  & 0 & 5  & 5 & 1 \\
                      &                     & IMF$_4$        & 3 & 10 & 5 & 9  & 5 & 1 \\
                      &                     & IMF$_5$        & 5 & 1  & 5 & 5  & 5 & 1 \\
                      &                     & Non-decomposed & 5 & 20 & 0 & 5  & 4 & 1 \\ \hline
\end{tabular}%
}
\end{table}
\end{document}